\documentclass[fleqn,usenatbib]{mnras}

\usepackage{newtxtext,newtxmath}

\usepackage[T1]{fontenc}

\DeclareRobustCommand{\VAN}[3]{#2}
\let\VANthebibliography\thebibliography
\def\thebibliography{\DeclareRobustCommand{\VAN}[3]{##3}\VANthebibliography}

\usepackage{graphicx}	
\usepackage{amsmath}	
\usepackage{soul}
\usepackage{array}

\newcommand{\dsec}{$\rlap{.}^s$}

\title[The magnetic field of NGC 253]{JCMT POL-2 observations of NGC 253: evidence for large-scale reordering of magnetic fields in giant molecular cloud complexes}

\author[K. Pattle \& W. K. Gear]{
Kate Pattle$^{1}$\thanks{E-mail: k.pattle@ucl.ac.uk (KP)}
and Walter K. Gear$^{2}$
\\
$^{1}$Department of Physics and Astronomy, University College London, Gower Street, London WC1E 6BT, United Kingdom\\
$^{2}$Centre for Astronomy, School of Natural Sciences, University of Galway, University Road, Galway H91 TK33, Ireland
}

\date{Accepted 2026 November 18. Received 2026 November 11; in original form 2026 August 10}

\pubyear{2026}

\begin{document}
\label{firstpage}
\pagerange{\pageref{firstpage}--\pageref{lastpage}}
\maketitle

\begin{abstract}
In this paper we present observations of the magnetic field of the starburst galaxy NGC 253 made in polarized 850\,$\mu$m dust emission using the POL-2 polarimeter on the James Clerk Maxwell Telescope (JCMT).  While the magnetic field typically traces the galactic spiral structure, we see significant deviations in field direction associated with the presence of stellar and/or supernova feedback, with the mean plane-of-sky magnetic field direction offset $24^{\circ}$ from the galactic plane.  Polarization fraction is anticorrelated with total 850\,$\mu$m intensity, with a power-law index of $0.72\pm 0.03$. Using the Fasano-Franceschini test, we find that the positions of POL-2 detections and radio sources in the spiral arms are inconsistent with one another, indicative of depolarization resulting from magnetic fields being tangled on scales smaller than the 240\,pc JCMT 850\,$\mu$m beam in sites of star formation and feedback.  Moreover, in both the starburst superwind and in one star-forming clump in the spiral arms, we observe a magnetic field direction consistent with being directed out of the plane of the galaxy.  We thus hypothesise that magnetic fields in the giant molecular clouds (GMCs) of NGC 253 are undergoing an evolutionary sequence from a relatively ordered field parallel to the spiral arms in the pre-star-formation gas, to a complex field caused by the interplay of turbulence, gravity and small-scale feedback in relatively young GMCs, to a field that is reordered out of the plane of the galaxy as large-scale H\textsc{ii} region expansion reshapes the gas and dust of galactic spiral arms.
\end{abstract}




\section{Introduction}

\begin{figure*}
    \centering
    \includegraphics[width=\textwidth]{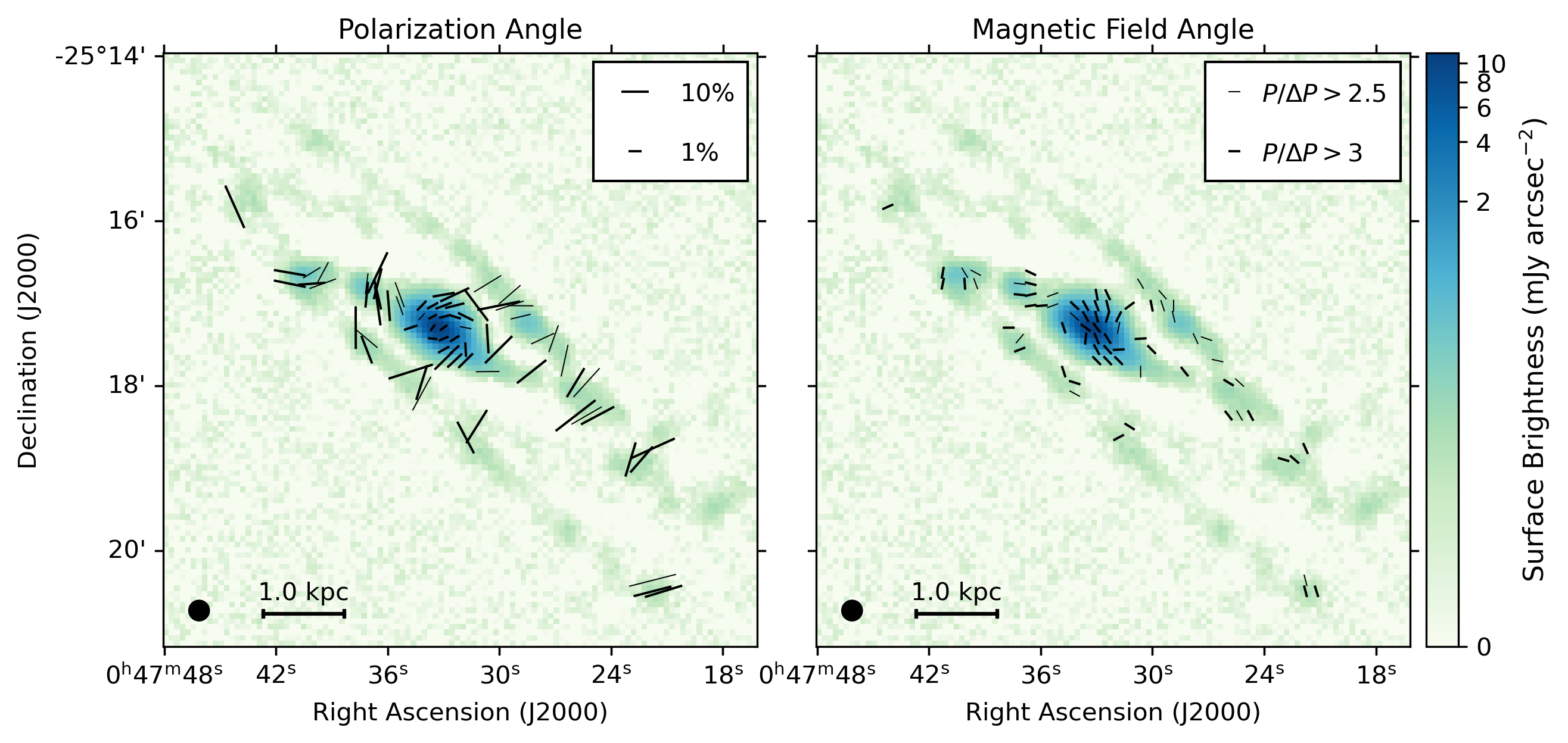}
    \caption{Our JCMT POL-2 observations of NGC 253.  Polarization and magnetic field vectors are overplotted on our POL-2 850\,$\mu$m Stokes $I$ (total intensity) image, the colour scale for which is shown to the right of the images. The Stokes $I$ images are shown on the default 4-arcsec pixel grid, with vectors overplotted on the 8-arcsec pixel grid used for analysis. Left: line segments are POL-2 850\,$\mu$m polarization vectors, with lengths proportional to the cube root of their polarization fraction, as shown in the key in the upper right of the panel. Right: line segments are POL-2 850\,$\mu$m magnetic field vectors (polarization vectors rotated by 90$^{\circ}$), plotted with constant length.  In both panels, the weight of the vectors indicates SNR; thick vectors have $\delta\theta < 9.6^{\circ}$ ($p/\delta p > 3$), while thin vectors have $ 9.6^{\circ} <\delta\theta < 11.5^{\circ}$ ($3 > p/\delta p > 2.5$).}
    \label{fig:ngc253}
\end{figure*}

Magnetic fields thread the interstellar media (ISM) of galaxies on all size scales, but their dynamic importance, and their role in regulating star formation, remains unclear \citep[e.g.][]{pattle2023a}.  Magnetic fields in spiral galaxies appear to be maintained by large-scale dynamo action \citep[e.g.][]{rieder2016, rieder2017, brandenburg2023}, with their directions generally being correlated with those of the spiral arms on galactic scales \citep[e.g.][]{scarrott1987,jones2020}.  However, on sub-kpc scales, magnetic field structures are more complex.  Magnetic fields in star-forming molecular clouds are typically oriented either parallel or perpendicular to local density structure \citep{soler2013}.  Moreover, star formation feedback appears to reshape magnetic field structures on a wide range of size scales, from individual protostellar outflows and H\textsc{ii} regions \citep[e.g.][]{karoly2023,khan2024}, through entire giant molecular clouds (GMCs; \citealt{lopezrodriguez2023,belfiori2025}), to wholesale rearrangement of a galactic dynamo field by a nuclear starburst galaxy \citep{thompson2006,adebahr2017,jones2019,jones2020,pattle2021a, lopezrodriguez2021}. 

NGC 253, also known as the Sculptor Galaxy, is a nearby, highly inclined ($74^{\circ}$; \citealt{iodice2014}), SAB(s)c barred spiral galaxy \citep{devaucouleurs1991}, the distance to which is generally found to be $3.5$ Mpc \citep{rekola2005,okamoto2024,newman2024}.  NGC 253 houses a nuclear starburst region, with a star formation rate of approximately $2.8$\,M$_{\odot}$\,yr$^{-1}$ \citep{ott2005}, which drives a molecular outflow with an outflow rate $>3\,$M$_{\odot}$\,yr$^{-1}$ \citep{bolatto2013}, as well as a superwind which has been studied across the electromagnetic spectrum, from radio to X-ray wavelengths \citep[][and refs. therein]{cronin2025}.  This starburst was likely triggered by a relatively recent encounter with a dwarf companion \citep{davidge2010}.  NGC 253 is also home to a number of star-forming GMC complexes located in its spiral arms \citep[e.g.][]{beslic2024}. 
We here present 850\,$\mu$m dust polarization observations at 240\,pc resolution of this archetypal starburst galaxy, and with them investigate the effects of significant stellar feedback on a galaxy-scale magnetic field.

The James Clerk Maxwell Telescope (JCMT) Submillimetre Common-User Bolometer Array 2 (SCUBA-2) camera is sensitive to thermal emission from interstellar dust grains \citep{holland2013}, preferentially tracing the coldest and densest molecular gas associated with star formation \citep{wardthompson2016}. 
SCUBA-2 850$\mu$m emission in nearby galaxies is well-correlated with star formation rate tracers \citep{pattle2023}, indicating that SCUBA-2 preferentially traces gas associated with star formation on galactic scales.

The major axes of interstellar dust grains are preferentially aligned perpendicular to the local ISM magnetic field direction throughout the large majority of the ISM \citep{andersson2015}, and so the preferential direction of polarization of thermal dust emission is perpendicular to the local plane-of-sky magnetic field direction.  The JCMT POL-2 polarimeter \citep{friberg2016}, operating in conjunction with SCUBA-2, has previously been used to map the magnetic field of the dense gas of the nearby nuclear starburst galaxy M82 \citep{pattle2021a}: these observations, preferentially tracing cold and dense gas in the plane of the galaxy, showed that in the central starburst region, the magnetic field had been fully rearranged to a poloidal configuration, tracing the direction of the galactic `superwind' driven by supernova feedback.  This behaviour was also seen in the warmer dust traced by the Stratospheric Observatory for Far-Infrared Astronomy (SOFIA; \citealt{jones2019}; \citealt{lopezrodriguez2021}).

NGC 253 has been observed in far-infrared polarized light by SOFIA with observations at 53\,$\mu$m, 89\,$\mu$m and 154\,$\mu$m \citep{jones2019,lopezrodriguez2022,borlaff2023}.  \citet{borlaff2023} note that the magnetic field of NGC253 is largely parallel to the galactic disc, with some field twisting in the central kpc of the galaxy, which they hypothesise arises from the galactic outflow associated with the central starburst.  \citet{jones2019} identify a similarly planar magnetic field geometry in NGC 253, but also note a weak indication of a vertical geometry above and below the central starburst region.

The magnetic field of the central starburst in NGC 253 was imaged at high resolution (0$\farcs$3; 5\,pc) at 890\,$\mu$m using the Atacama Large Millimeter Array (ALMA) by \citet{lopezrodriguez2023}, who observe two magnetic field components: in the star-forming clusters, the field is perpendicular to the galactic disc, while in the lower column density material between these clusters, the field remains parallel to the galactic disc, as is seen on larger scales.  These ALMA observations were further analysed by \citet{belfiori2025}, who demonstrated that the 890\,$\mu$m data trace only dust emission, while ALMA 2.1\,mm observations, which arise from a mixture of dust emission, free-free and synchrotron components, trace the magnetic field at the base of the galactic outflow.  \citet{heesen2011}, observing NGC 253 in 3cm and 6cm polarized emission, observed radio filaments above and below the galactic centre which they interpreted as the walls of the outflow cone from the nuclear starburst.  They further identified a helical magnetic field in the outflow, and found that the outflow is collimated by the magnetic field.

In this paper we investigate the magnetic field in the cold molecular ISM of NGC 253, as traced by JCMT POL-2 850\,$\mu$m dust polarization observations.  The paper is laid out as follows: Section~\ref{sec:obs} presents our observations.  Section~\ref{sec:results} presents our results, which are discussed in Section~\ref{sec:discussion}.  Our conclusions are presented in Section~\ref{sec:conclusions}.

\section{Observations}
\label{sec:obs}

NGC 253 was observed 17 times between 2021 November 29 and 2023 January 11 using the POL-2 polarimeter \citep{friberg2016} mounted on the Submillimetre Common-User Bolometer Array 2 (SCUBA-2; \citealt{holland2013}) on the James Clerk Maxwell Telescope (JCMT).  The data were taken in JCMT Weather Band 1 ($\tau_{225\,{\rm GHz}}<0.05$), under project codes M21BP024 and M22BP006.  Each observation consisted of a 31-minute POL-2 DAISY scan pattern, for a total on-source time of 8.8 hr.

The data were reduced using the $pol2map$\footnote{\url{http://starlink.eao.hawaii.edu/docs/sun258.htx/sun258ss73.html}} script, from the \textsc{Smurf} package in the \textit{Starlink} software suite \citep{chapin2013}.  See \citet{pattle2021} for a detailed description of the POL-2 data reduction process. Instrumental polarisation (IP) was corrected for using the `August 2019' IP model\footnote{\url{https://www.eaobservatory.org/jcmt/2019/08/new-ip-models-for-pol2-data/}}.  The 850$\mu$m data were calibrated using a flux conversion factor (FCF) of 2795 mJy\,arcsec$^{-2}$\,pW$^{-1}$ using the post-2018 June 30 SCUBA-2 FCF of 2070 mJy\,arcsec$^{-2}$\,pW$^{-1}$ \citep{mairs2021} multiplied by a factor of 1.35 to account for additional losses in POL-2 \citep{friberg2016}.  POL-2 observes simultaneously at 850$\mu$m and 450$\mu$m, but we consider only the 850$\mu$m observations in this work.  The resolution of the JCMT at 850$\mu$m is 14.1$^{\prime\prime}$ \citep{dempsey2013}, equivalent to a linear resolution of 240\,pc at a distance of 3.5 Mpc.

The average RMS noise in the central 3 arcmin of the map, over which exposure time is approximately constant \citep{friberg2016}, on the default 4-arcsec pixel scale used in POL-2 data reduction is 4.0 mJy\,beam$^{-1}$ (Stokes $I$), 3.7 mJy\,beam$^{-1}$ (Stokes $Q$), and 3.8 mJy\,beam$^{-1}$ (Stokes $U$).  In keeping with common practice, we binned our output vector catalogue to 8-arcsec (approximately Nyquist-sampled) pixels.  The average RMS noise in Stokes $Q$ and $U$ and $I$ in the central 3 arcmin of the map on 8-arcsec pixels is 1.3 mJy\,beam$^{-1}$.

The observed polarized intensity is given by
\begin{equation}
    PI^{\prime} = \sqrt{Q^{2} + U^{2}}.
\end{equation}
We debiased this quantity using the modified asymptotic estimator \citep{plaszczynski2014,montier2015}:
\begin{equation}
    PI = PI^{\prime} - \frac{1}{2}\frac{\sigma^{2}}{PI^{\prime}}\left(1-e^{-\left(\frac{PI^{\prime}}{\sigma}\right)^{2}}\right),
\end{equation}
where $\sigma^{2}$ is the weighted mean of the variances {$\sigma_{Q}^{2}$ and $\sigma_{U}^{2}$},
\begin{equation}
    \sigma^{2} = \frac{Q^{2}\sigma_{Q}^{2} + U^{2}\sigma_{U}^{2}}{Q^{2} + U^{2}},
\end{equation}
calculated on a pixel-by-pixel basis. Debiased polarisation fraction is given by $p = PI/I$.

Polarization angle is given by
\begin{equation}
    \theta_{p} = \frac{1}{2}\arctan\left(\frac{U}{Q}\right).
\end{equation}
We note that the polarisation angles which we detect are not true vectors, as they occupy a range in angle $0-180^{\circ}$.  We nonetheless refer to our measurements as vectors for convenience, in keeping with the general convention in the field.

Throughout this work we assume that dust grains are aligned with their major axis perpendicular to the magnetic field direction \citep[e.g.][]{andersson2015}, and so that the plane-of-sky magnetic field direction can be inferred by rotating $\theta_{p}$ by 90$^{\circ}$.

\begin{figure}
    \centering
    \includegraphics[width=0.47\textwidth]{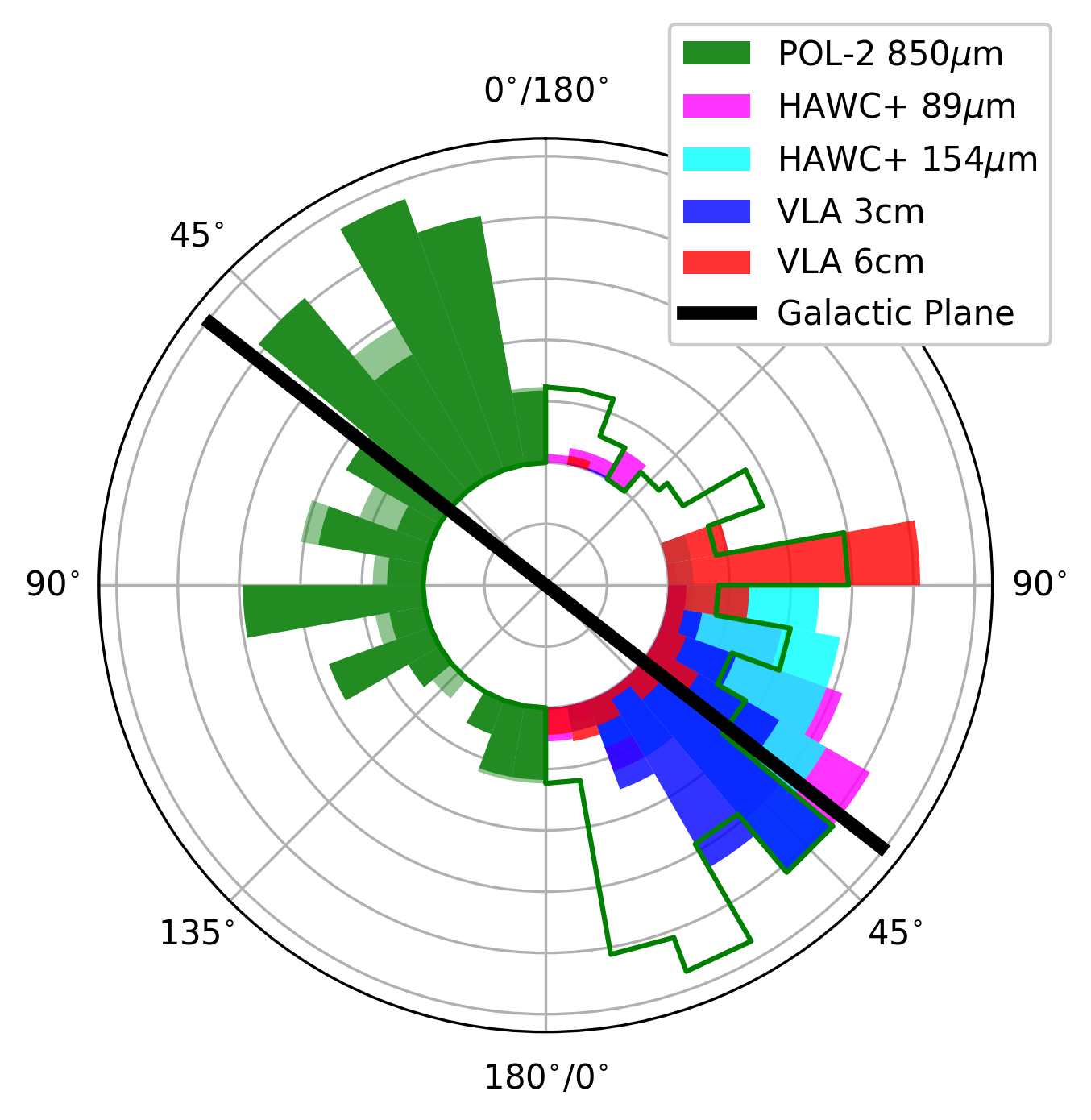}
    \caption{A polar histogram of magnetic field angles measured in NGC 253.  The left-hand semicircle shows POL-2 data: solid green indicates $p/\delta p > 3$, while transparent green indicates $p/\delta p > 2.5$.  HAWC+ and VLA magnetic field angles are shown in the opposite semicircle, with HAWC+ Band C shown in magenta, HAWC+ Band D shown in cyan, VLA 3\,cm radio in blue and 6\,cm radio in red.  The outline of the POL-2 histogram is shown in green, for comparison.  The position angle of the galactic plane (the major axis of the nuclear ring and outer ring) is shown as a black line.  Due to the $\pm 180^{\circ}$ ambiguity on polarization vector measurements, opposite angles on the plot agree.  In Cartesian space, the histogram areas are normalized to 1. The projection of the histograms on to a circle means that their areas are distorted.}
    \label{fig:polar}
\end{figure}

\section{Results}
\label{sec:results}

We show the 850\,$\mu$m total intensity emission and dust polarization vectors that we measure in NGC 253 in Figure~\ref{fig:ngc253}.  We show all vectors with $\delta\theta < 11.5^{\circ}$ (equivalent to $p/\delta p > 2.5$; \citealt{serkowski1962}) and $I/\delta I > 5$. Vector weights in Figure~\ref{fig:ngc253} show signal-to-noise ratio in terms of $p/\delta p$. It can be seen that the position angles of vectors with $\delta\theta < 11.5^{\circ}$ agree well with those with $\delta\theta < 9.6^{\circ}$ (equivalent to $p/dp > 3$; \citealt{serkowski1962}), and so we include them in our analysis.  We choose to define our uncertainties using uncertainty in polarization angle rather than polarization fraction since polarization angle is the more statistically robust quantity \citep{naghizadehkhouei1993}, and since polarization angle is the quantity with which we are primarily concerned.

\subsection{Magnetic field morphology}

The distribution of magnetic field angles in NGC 253 is shown in Figures~\ref{fig:ngc253} and~\ref{fig:polar}.  The mean magnetic field angle that we measure with POL-2 is $29^{\circ}$ E of N, with substantial variations.  It can be seen that the magnetic field in NGC 253 broadly follows the spiral structure of the galaxy, albeit with a significant excursion above the galactic nucleus.  We also see some deviations from the direction of the spiral arms at particular positions within the arms, which are discussed in detail in Section~\ref{sec:discussion_centre}.

Figure~\ref{fig:polar} also shows the distribution of magnetic field angles measured in NGC253 in the far-infrared with the High-resolution Airborne Wide-band Camera (HAWC+) on the Stratospheric Observatory for Infrared Astronomy (SOFIA) as part of the Survey of Extragalactic Magnetism with SOFIA (SALSA) \citep{jones2019, lopezrodriguez2022}, and in the radio using the Very Large Array (VLA) \citep{heesen2011}.  Figure~\ref{fig:polar} shows that while these different wavelengths appear to observe a qualitatively similar magnetic field direction to our 850\,$\mu$m data, all broadly parallel to the plane of the galaxy, there are systematic offsets between the different wavelengths.  Notably, the POL-2 data appears to trace a magnetic field which not fully aligned with the plane of the galaxy, as we discuss in detail below.  The HAWC+ and VLA data sets are discussed in detail in Appendix~\ref{sec:appendix}; relevant details of the 89\,$\mu$m data are discussed in Section~\ref{sec:discussion_reorg_stats}.

\subsection{Model of spiral structure}

\begin{figure}
    \centering
    \includegraphics[width=0.47\textwidth]{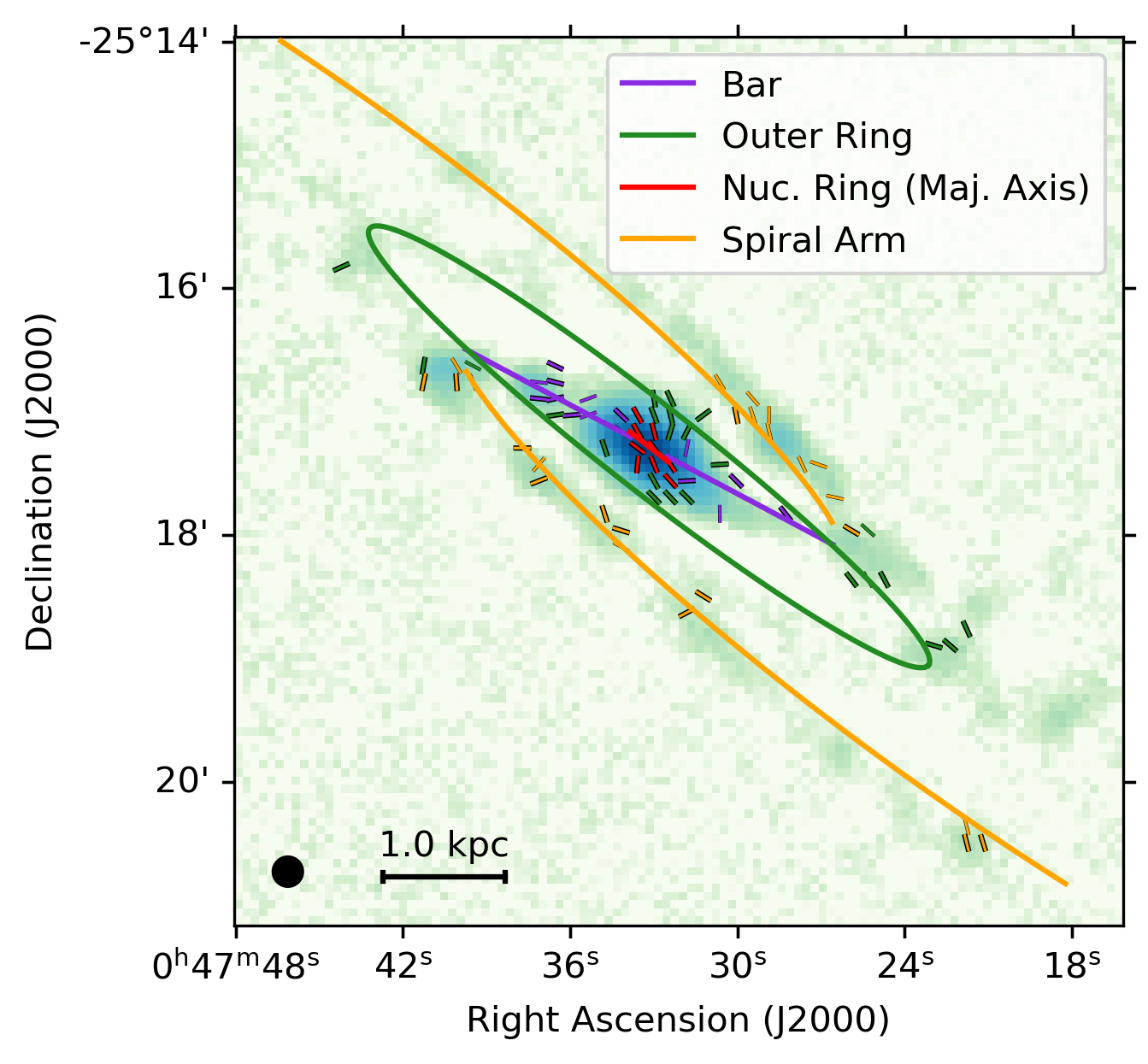}
    \caption{Our empirical spiral structure model for NGC253, overlaid on POL-2 850\,$\mu$m Stokes $I$ emission.  POL-2 magnetic field vectors are colour coded by the spiral feature to which they are most proximate.}
    \label{fig:spiral}
\end{figure}

We created a simple empirical model of the spiral structure of NGC 253, as shown in Figure~\ref{fig:spiral}.  This model consists of the accepted components of NGC 253 \citep{iodice2014}: the nuclear ring (here represented by the position angle of its major axis), the bar, the outer ring, and a pair of $m=1$ spiral arms.

Both the nuclear ring and the outer ring have position angles of 52$^{\circ}$ E of N, following \citep{iodice2014}.  We use an empirical position angle for the bar of 62$^{\circ}$.  \citet{iodice2014} give a position angle for the bar of 72$^{\circ}$, but we found that this was not a suitable choice for our data, possibly due to NGC253 being a highly-inclined galaxy: \citet{iodice2014} measured the structure of NGC 253 in near-infrared light, and so may be preferentially tracing the near side of the bar, leading to a higher observed position angle, while our optically-thin 850$\mu$m observations trace the full column of material associated with the bar.

We represent the $m=1$ spiral arms using a lognormal model, $r=A\exp(B\phi)$, taking $A=156.25$\,pc (12.5\% of a 1250\,pc bar length), $B=\pi/6$ radians, and the rotation angle of the spiral structure to be 52$^{\circ}$, for $\phi$ over the range $1.3\pi \leq \phi \leq 1.8\pi$ radians.  It must be emphasised that this is an approximate empirical model only, and is suitable only over the range of angles specified.

\subsection{Clumpfind}

\begin{table*}
\centering
\begin{tabular}{c cc cc @{\extracolsep{4pt}} cc c}
\hline
 & & & \multicolumn{2}{c}{Flux Density} & \multicolumn{2}{c}{Position Angle} & U2000 \\ \cline{4-5} \cline{6-7} 
Clump ID & R.A. & Dec. & Peak & Total & B-Field & Spiral & ID \\ \cline{2-3} \cline{6-7} 
 & \multicolumn{2}{c}{(J2000)} & mJy/beam & mJy/arcsec$^{2}$ & \multicolumn{2}{c}{($^{\circ}$ E of N)} & \\ 
\hline
1 & 0$^{h}$47$^{m}$33\dsec2 & --25$^{\circ}$17$^{\prime}$19$^{\prime\prime}$ & 2989$\pm$179 & 5788$\pm$347 & 29$\pm$36 & 54.7$\pm$0.3 & -- \\ 
2 & 0$^{h}$47$^{m}$28\dsec5 & --25$^{\circ}$17$^{\prime}$14$^{\prime\prime}$ & 127$\pm$9 & 176$\pm$11 & 26$\pm$23 & 41.5$\pm$0.9 & 8, 7, 6, 4 \\ 
3 & 0$^{h}$47$^{m}$37\dsec1 & --25$^{\circ}$16$^{\prime}$50$^{\prime\prime}$ & 145$\pm$10 & 137$\pm$8 & 85$\pm$13 & 60.9$\pm$0.1 & 18 \\ 
4 & 0$^{h}$47$^{m}$40\dsec3 & --25$^{\circ}$16$^{\prime}$42$^{\prime\prime}$ & 125$\pm$8 & 197$\pm$12 & 13$\pm$26 & 47.2$\pm$0.7 & 20, 21, 19 \\ 
5 & 0$^{h}$47$^{m}$25\dsec4 & --25$^{\circ}$18$^{\prime}$09$^{\prime\prime}$ & 50$\pm$5 & 105$\pm$7 & 46$\pm$12 & 47.2$\pm$0.6 & -- \\ 
6 & 0$^{h}$47$^{m}$37\dsec1 & --25$^{\circ}$17$^{\prime}$31$^{\prime\prime}$ & 58$\pm$5 & 61$\pm$4 & 113$\pm$21 & 43.0$\pm$1.0 & 17 \\ 
7 & 0$^{h}$47$^{m}$26\dsec8 & --25$^{\circ}$17$^{\prime}$34$^{\prime\prime}$ & 82$\pm$6 & 38$\pm$3 & 78 & 34.2$\pm$0.4 & 2, 1 \\ 
8 & 0$^{h}$47$^{m}$30\dsec4 & --25$^{\circ}$16$^{\prime}$44$^{\prime\prime}$ & 59$\pm$5 & 40$\pm$3 & 31 & 45.2$\pm$0.5 & 10, 11, 12 \\ 
9 & 0$^{h}$47$^{m}$18\dsec4 & --25$^{\circ}$19$^{\prime}$28$^{\prime\prime}$ & 57$\pm$5 & 40$\pm$3 & -- & -- & -- \\ 
10 & 0$^{h}$47$^{m}$30\dsec6 & --25$^{\circ}$18$^{\prime}$54$^{\prime\prime}$ & 43$\pm$5 & 36$\pm$3 & -- & 50.5$\pm$0.5 & -- \\ 
11 & 0$^{h}$47$^{m}$22\dsec3 & --25$^{\circ}$18$^{\prime}$59$^{\prime\prime}$ & 55$\pm$5 & 37$\pm$3 & 37$\pm$13 & -- & -- \\ 
12 & 0$^{h}$47$^{m}$21\dsec7 & --25$^{\circ}$20$^{\prime}$30$^{\prime\prime}$ & 66$\pm$6 & 28$\pm$2 & 15$\pm$1 & 56.0$\pm$0.1 & -- \\ 
13 & 0$^{h}$47$^{m}$23\dsec4 & --25$^{\circ}$18$^{\prime}$58$^{\prime\prime}$ & 74$\pm$6 & 23$\pm$2 & 73 & 15.8$\pm$1.3 & -- \\ 
14 & 0$^{h}$47$^{m}$43\dsec4 & --25$^{\circ}$15$^{\prime}$44$^{\prime\prime}$ & 43$\pm$5 & 40$\pm$3 & 114 & 21.8$\pm$1.4 & 22, 25 \\ 
15 & 0$^{h}$47$^{m}$34\dsec6 & --25$^{\circ}$18$^{\prime}$00$^{\prime\prime}$ & 105$\pm$7 & 23$\pm$2 & 67$\pm$6 & 46.8$\pm$0.3 & -- \\ 
16 & 0$^{h}$47$^{m}$31\dsec8 & --25$^{\circ}$18$^{\prime}$40$^{\prime\prime}$ & 60$\pm$5 & 20$\pm$2 & 118 & 49.1$\pm$0.3 & -- \\ 
17 & 0$^{h}$47$^{m}$35\dsec5 & --25$^{\circ}$17$^{\prime}$49$^{\prime\prime}$ & 62$\pm$5 & 19$\pm$2 & 18 & 45.5$\pm$0.5 & -- \\ 
18 & 0$^{h}$47$^{m}$31\dsec8 & --25$^{\circ}$16$^{\prime}$23$^{\prime\prime}$ & 53$\pm$5 & 20$\pm$2 & -- & -- & 13 \\ 
19 & 0$^{h}$47$^{m}$39\dsec9 & --25$^{\circ}$15$^{\prime}$01$^{\prime\prime}$ & 59$\pm$5 & 18$\pm$2 & -- & 52.9$\pm$0.3 & 23 \\ 
20 & 0$^{h}$47$^{m}$20\dsec9 & --25$^{\circ}$19$^{\prime}$25$^{\prime\prime}$ & 105$\pm$7 & 10$\pm$1 & -- & -- & -- \\ 
\hline
\end{tabular}
\caption{Structures identified in 850\,$\mu$m Stokes $I$ emission using \textit{Clumpfind}.  The final column lists \citet{ulvestad2000} sources that lie within the boundary of the clump, ordered by proximity to the clump's centre.}
\label{tab:clumpfind}
\end{table*}

\begin{figure}
    \centering
    \includegraphics[width=0.47\textwidth]{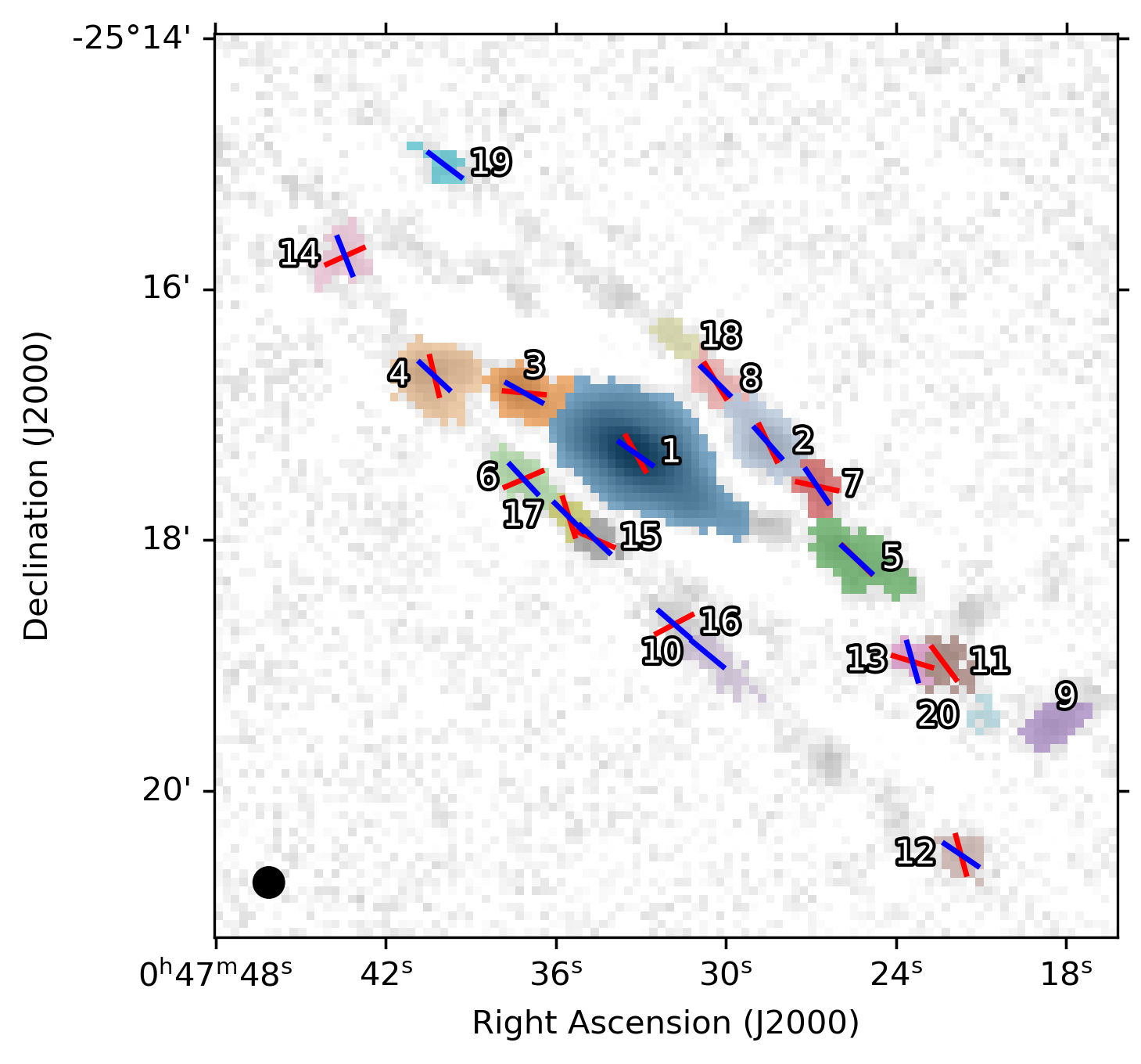}
    \caption{Average magnetic field angle (red) and average spiral arm angle (blue) overplotted on coloured pixels demarcating \textit{Clumpfind} clumps, which are numbered as in Table~\ref{tab:clumpfind}.  The background image shows POL-2 850$\mu$m Stokes $I$ emission.}
    \label{fig:clumps}
\end{figure}

To aid comparison between the magnetic field and spiral arm directions in NGC 253, we identified emission peaks in the POL-2 Stokes $I$ emission, and determined the average magnetic field direction and spiral arm angle in each clump.  To do so, we used the implementation of the \textit{Clumpfind} algorithm \citep{williams1994} in the \textit{Starlink} \textsc{Cupid} package \citep{currie2014}.  \textit{Clumpfind} segments data using a friends-of-friends algorithm, down to a specified contour level in the data; here, twice the RMS noise in Stokes $I$ (i.e. 8.0 mJy/beam).  We identified 20 clumps in the 850 $\mu$m Stokes $I$ emission, which are shown in Figure~\ref{fig:clumps} and listed in Table~\ref{tab:clumpfind}.

We calculated the mean magnetic field angle and spiral arm direction in each clump using the \textit{scipy} circular mean function.  For clumps with more than one vector, we calculated the spread in angles using the \textit{scipy} circular variance function.  These mean angles, and their associated standard deviations, are listed in Table~\ref{tab:clumpfind}, and the mean vectors are shown on Figure~\ref{fig:clumps}.

\subsection{Relationship between polarization properties and star formation rate}

\begin{figure*}
    \centering
    \includegraphics[width=0.75\textwidth]{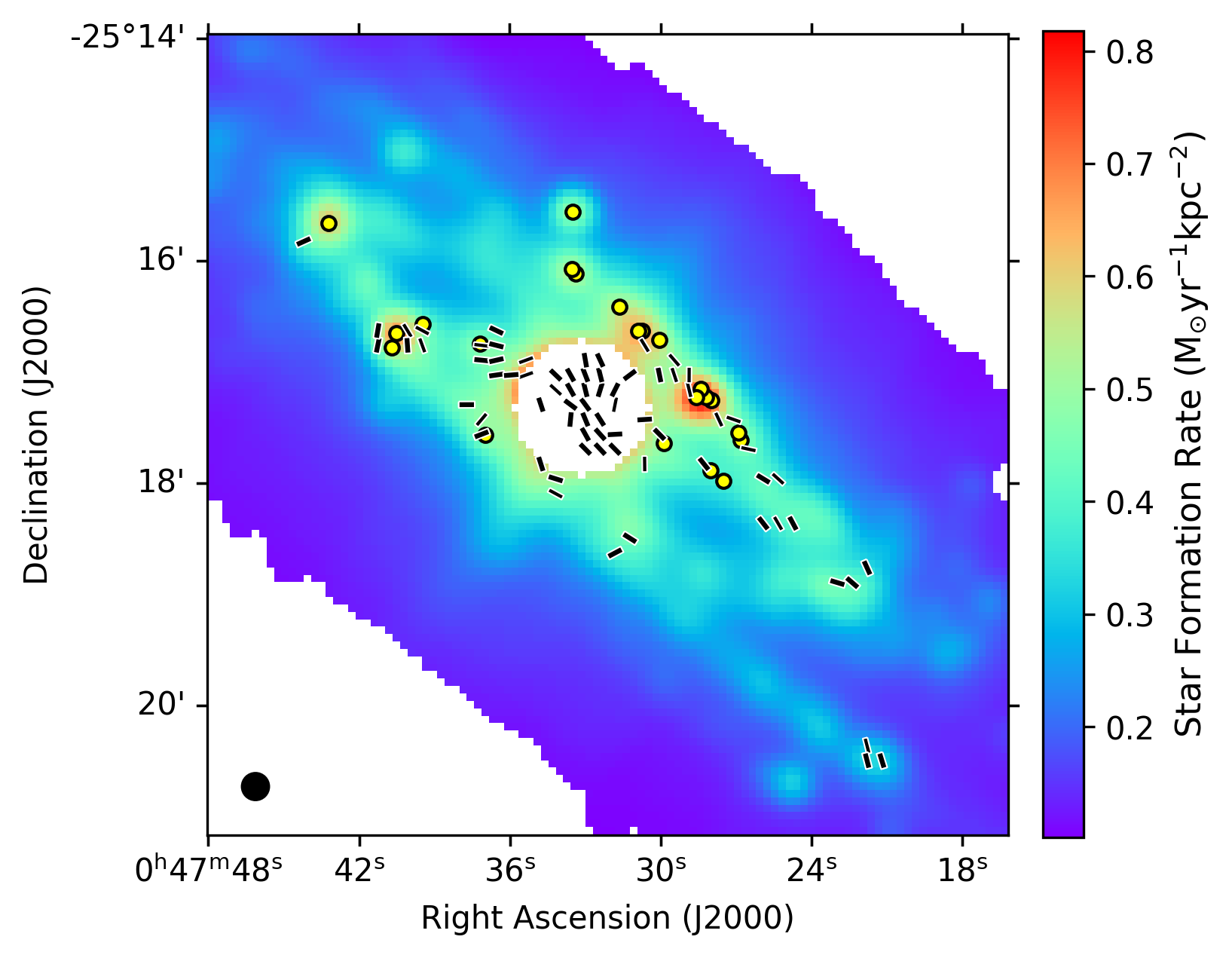}
    \caption{Star formation rate surface density in NGC 253, with POL-2 vectors overplotted as black lines and \citet{ulvestad2000} radio sources overplotted as yellow circles.  The centre of NGC 253 is saturated in the \textit{Spitzer} MIPS 24$\mu$m channel, and so the central 36$^{\prime\prime}$ of the map is removed from this image.}
    \label{fig:sfr}
\end{figure*}

Figure~\ref{fig:sfr} shows the surface density of star formation rate in NGC253, which we measured using the method described by \citet{leroy2008} and \citet{ford2013}.  We measured unobscured star formation using GALEX far-ultraviolet (FUV) emission ($I_{\textsc{fuv}}$) and obscured star formation using \textit{Spitzer} Space Telescope 24$\mu$m emission ($I_{24}$).  The older stellar population was accounted for using \textit{Spitzer} 3.6$\mu$m emission ($I_{3.6}$).  All of the maps used were taken from the Dustpedia database \citep{clark2018}.  We smoothed all of the data sets to $14\farcs1$ resolution, and gridded them to the POL-2 Stokes $I$ map.

Any pixels with significant GALEX near-ultraviolet (NUV) emission, i.e. where $I_{\textsc{nuv}}/I_{\textsc{fuv}} > 15$, were excluded from the analysis, to avoid contamination by foreground stars \citep{leroy2008}.
We then estimated the FUV flux density associated with star formation ($I_{\textsc{fuv,sf}}$) using the relation
\begin{equation}
    I_{\textsc{fuv,sf}} = I_{\textsc{fuv}} - \alpha_{\textsc{fuv}}I_{3.6},
\end{equation}
taking $\alpha_{\textsf{fuv}}=3\times10^{-3}$, and the 24$\mu$m flux density associated with star formation ($I_{24,\textsc{sf}}$) using the relation
\begin{equation}
    I_{24,\textsc{sf}} = I_{24} - \alpha_{24}I_{3.6},
\end{equation}
taking $\alpha_{24}=0.1$ \citep{leroy2008}.
We then estimated the surface density of star formation to be
\begin{equation}
    \Sigma_{\textsc{sfr}} = 8.1\times10^{-2}I_{\textsc{fuv,sf}} + 3.2\times10^{-3}I_{24,\textsc{sf}}.
    \label{eq:ssfr}
\end{equation}

The \textit{Spitzer} 24$\mu$m channel was saturated in the nuclear region of NGC 253.  We thus excluded an area of radius 36$^{\prime\prime}$, centred on the galactic nucleus.  Low-level radial features arising from the 24$\mu$m diffraction spikes remain visible in our map of star formation rate surface density, but these do not significantly affect our results.

\subsection{Relationship between polarisation fraction and intensity}
\label{sec:pI}

\begin{figure}
    \centering
    \includegraphics[width=0.47\textwidth]{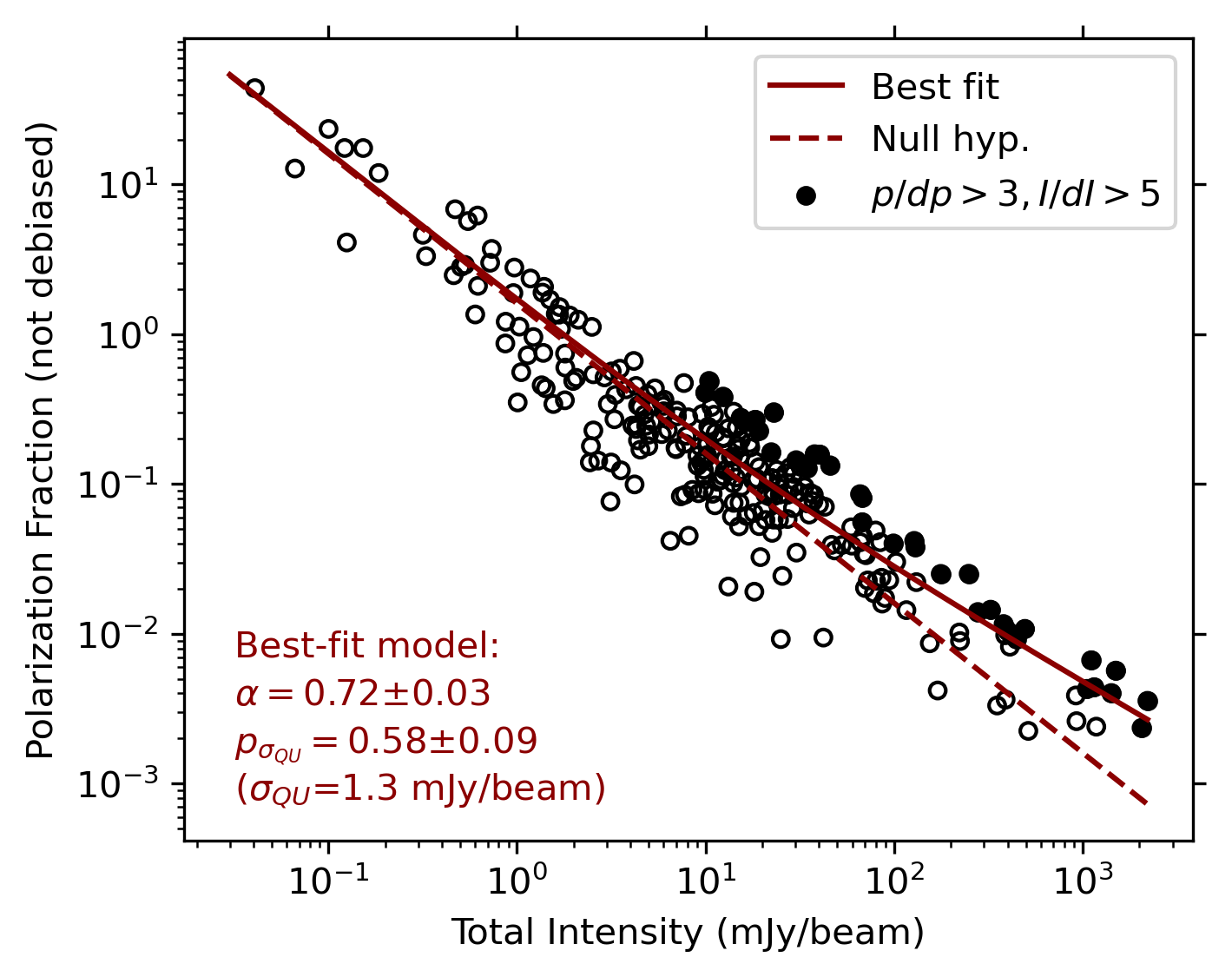}
    \caption{Non-debiased polarization fraction as a function of Stokes $I$ intensity, fitted with a single-power-law distribution and a Ricean noise model, as described in the text.  All data points in the central 3-arcmin-diameter region of the image are shown and fitted; those at $p/dp > 3$ are shaded in black.  The best-fit model, with a power-law index $\alpha=0.72\pm0.03$, is shown as a solid maroon line.  The behaviour in the absence of true polarised signal, $p^{\prime} = \sqrt{\pi/2}(I/\sigma_{QU})^{-1}$, is shown as a dashed maroon line.}
    \label{fig:p_vs_I}
\end{figure}

Polarized dust emission observations typically show a power-law dependence, $p\propto I^{-\alpha}$, where $0\leq \alpha \leq 1$ \citep{whittet2008,jones2015}.  A steeper index (higher $\alpha$) indicates either poorer grain alignment with respect to the magnetic field or more variation of the magnetic field direction along the line of sight (LOS): $\alpha = 0$ indicates that grains are consistently aligned throughout the LOS, while $\alpha = 1$ implies complete randomization of either grain alignment or magnetic field direction along the LOS \citep{pattle2019a}.  Given that the linear resolution of our observations is 240\,pc, we expect that a negative power-law index would arise from tangling of the magnetic field on scales smaller than the beam, rather than from loss of alignment of dust grains with respect to the magnetic field, as is seen on sub-parsec scales within dense star-forming cores within Milky Way star-forming regions \citep[e.g.,][]{alves2014}.

We measured $\alpha$ using the method described by \citet{pattle2019a}, in which we assume that the underlying relationship between $p$ and $I$ can be parametrized as
\begin{equation}
    p = p_{\sigma_{QU}}\left(\frac{I}{\sigma_{QU}}\right)^{-\alpha}
    \label{eq:polfrac}
\end{equation}
where $p_{\sigma_{QU}}$ is the polarization fraction at the RMS noise level of the data $\sigma_{QU}$, and $\alpha$ is a power-law index in the range $0 \leq \alpha \leq 1$.  We fitted the relationship between $I$ and observed non-debiased polarisation fraction $p^{\prime}$ with the mean of the Ricean distribution of observed values of $p^{\prime}$ which would arise from equation~\ref{eq:polfrac} in the presence of Gaussian RMS noise $\sigma_{QU}$ in Stokes $Q$ and $U$:
\begin{equation}
    p^{\prime}(I) = \sqrt{\frac{\pi}{2}}\left(\frac{I}{\sigma_{QU}}\right)^{-1}\mathcal{L}_{\frac{1}{2}}\left(-\frac{p_{\sigma_{QU}}^{2}}{2}\left(\frac{I}{\sigma_{QU}}\right)^{2(1-\alpha)}\right).
    \label{eq:rmfit}
\end{equation}
where $\mathcal{L}_{\frac{1}{2}}$ is a Laguerre polynomial of order $\frac{1}{2}$.  See \citet{pattle2019a} for a derivation of this result.  We restricted our data set to the central 3-arcminute diameter region over which exposure time, and so RMS noise, is approximately constant \citep{friberg2016}.

The relationship between $p^{\prime}$ and $I$ in NGC 253 is shown in Figure~\ref{fig:p_vs_I}.  By fitting Equation~\ref{eq:rmfit} to the data, we measure a best-fit index of $\alpha = 0.72\pm 0.03$.  This indicates that while the magnetic field generally remains sufficiently well-ordered for polarised dust emission to to provide information on the magnetic field direction on scales of a few hundred pc, there is an increasing randomization of dust grain orientation -- and thus, implicitly, of the magnetic field direction -- as surface brightness increases.  

We note that the $\alpha$ index is significantly steeper in NGC 253 than in M82, where $\alpha = 0.25\pm0.08$ \citep{pattle2021a}.  This suggests that the magnetic field has more small-scale complexity in NGC 253 than in M82.  This is in keeping with the magnetic field morphologies which we observe in the two galaxies: as discussed in Section~\ref{sec:discussion} below, in the centre of NGC 253, we appear to see contributions to the polarized emission both from dust grains tracing outflow-entrained fields, and from grains tracing fields following the major axis of the nuclear ring -- a behaviour that was also observed by \citet{lopezrodriguez2023} in the central starburst in ALMA observations.  Integration of these two broadly orthogonal components along the line of sight is likely to result in a cancellation of polarized signal, and so a steeper $\alpha$.  Conversely, in M82, the magnetic field in the central starburst appears to be fully poloidal and aligned with the outflow and superwind from the galaxy, without any indication of multiple components along the line of sight.

\subsubsection{Comparison with HAWC+ data}

\citet{lopezrodriguez2022}, analysing HAWC+ observations of NGC 253, investigated the power-law index of the relation between $p$ and total column density, $N_{\rm HI+H2}$, finding $\alpha$ values of $0.43^{+0.13}_{-0.12}$ at 154$\mu$m and $1.0^{+0.01}_{-0.01}$ 89\,$\mu$m.

Column density is only directly proportional to intensity in isothermal material \citep{hildebrand1983}, and the \citet{lopezrodriguez2022} analysis uses debiased $p$ values. Moreover, with the exception of a few 89\,$\mu$m vectors in the spiral arms, the HAWC+ data are largely restricted to the galactic centre and bar (see Figure~\ref{fig:hawc_data}), while our observations also trace the spiral arms.  Only qualitative comparison of our $\alpha$ values and those of \citet{lopezrodriguez2022} is thus possible.  However, we note that the index of 1 seen in 89\,$\mu$m emission, tracing hot dust, is indicative of near-total depolarization in feedback-heated regions.  The shallower index seen at 154$\mu$m, tracing material at temperatures more similar to our own 850\,$\mu$m data, suggests a more ordered field in the cooler material making up the bulk of the spiral arms and bar.

\subsubsection{Polarisation fraction in debiased data}

The analysis in the preceding section makes use of data that has not been statistically debiased, since this allows us to assume that the noise in $p^{\prime}$ is Ricean-distributed \citep{pattle2019a}.  We note that the polarization fractions that we measure in our debiased data include some unphysically high values of $>20$\,\% (cf. \citealt{planck2020}), as shown in Figure~\ref{fig:ngc253}.  Conversely, the maximum polarisation fractions in the two HAWC+ bands that we consider are $<5$\,\%, but these measurements are largely localised to the galactic centre, where we see the lowest polarization fractions in our 850\,$\mu$m data, typically also $<5$\,\%.  It is likely that our statistical debiasing of polarization fraction is incomplete in the faint, lower-SNR data in the spiral arms \citep{plaszczynski2014}; this is possibly exacerbated by the loss of extended flux in SCUBA-2/POL-2 observations \citep{sadavoy2013,pattle2023}.  However, we note that since the Stokes $Q$, $U$ and $I$ timestreams are all treated identically in POL-2 data reduction \citep{pattle2021}, any loss of extended flux should apply to polarized and total intensity equally.  Artificially high polarization fractions thus likely result from lowered SNR in both $PI$ and $I$, rather than loss of flux in $I$ alone.  We further note that we do not use debiased polarization fraction for any analysis in this work: our further analysis relies only on polarization angle measurements, which are significantly more statistically robust than are polarization fraction measurements \citep{naghizadehkhouei1993}.

\section{Discussion}
\label{sec:discussion}

We discuss separately the magnetic field that we observe in the vicinity of the galactic nuclear starburst and its associated outflow, and the magnetic fields that we observe in the spiral arms of the galaxy and their embedded GMCs.

\subsection{Galactic Centre}
\label{sec:discussion_centre}

\begin{figure}
    \centering
    \includegraphics[width=0.47\textwidth]{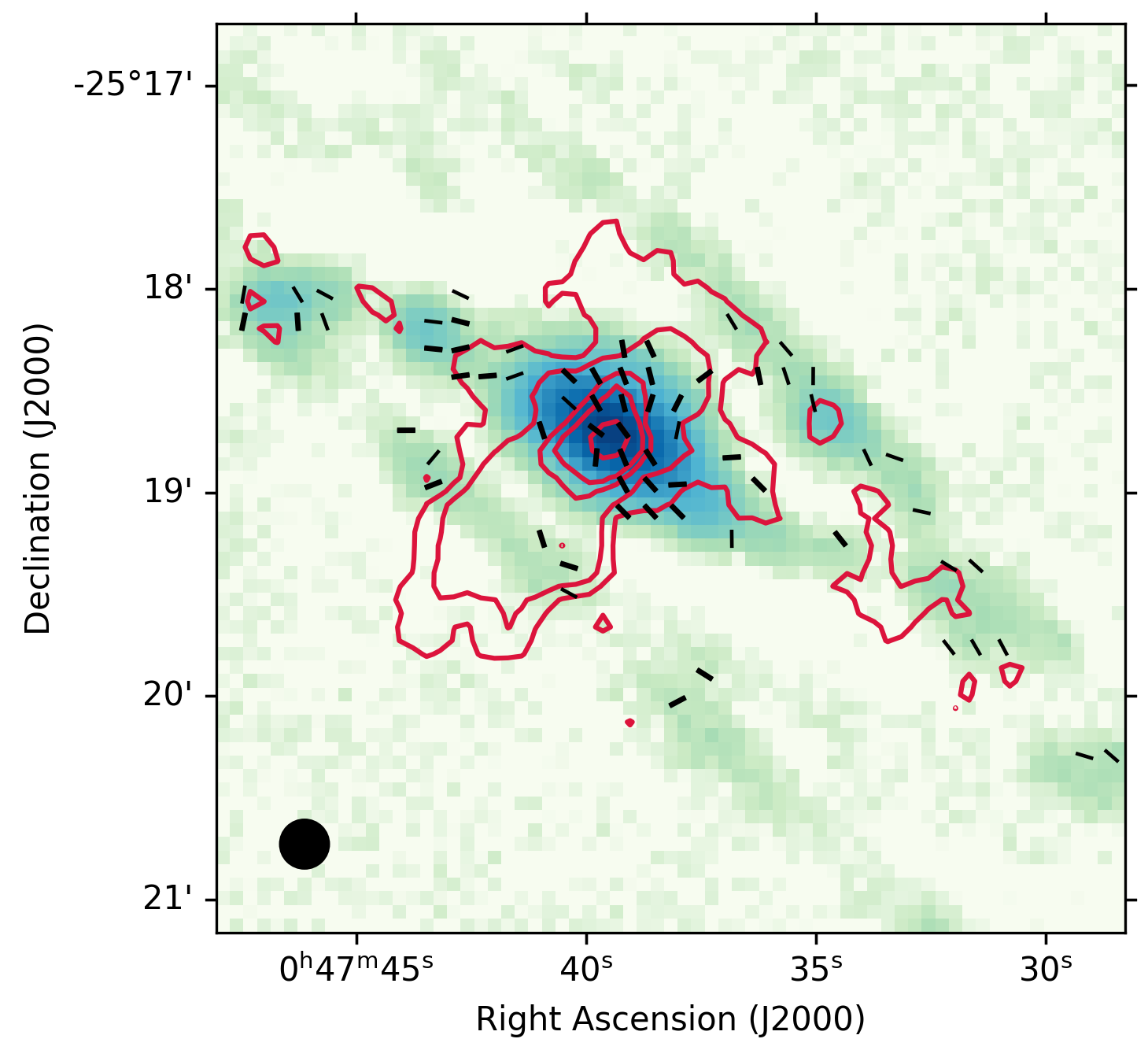}
    \caption{Contours show X-ray emission in NGC 253 from the Chandra X-ray Observatory \citep{lopez2023}, tracing the outflow from the nuclear starburst and intense star formation activity in the arms.  Contours show 0.005, 0.01, 0.05, 0.1, and 0.5$\times$ the peak X-ray emission value.  Background shows POL-2 850\,$\mu$m Stokes $I$ emission, with the same colour scaling as Figure~\ref{fig:ngc253}.}
    \label{fig:xray}
\end{figure}

Above the galactic nucleus in NGC253 the field appears to trace the galactic outflow, as shown in X-ray emission \citep{lopez2023} in Figure~\ref{fig:xray}, which entrains both dense gas and dust \citep{cronin2025}.  Similarly to the behaviour seen in 850\,$\mu$m polarization observations of M82 \citep{pattle2021a}, we appear to preferentially observe magnetic fields frozen into material that has been entrained in the starburst-driven outflow.  Interestingly, this apparent reordering of the magnetic field by the outflow is only visible in the red-shifted northern outflow lobe, which is behind the galactic disc.

The reordering of the field is less pronounced than in M82, perhaps due to the correspondingly weaker outflow from NGC 253: the kinetic energy of the NGC 253 outflow is $\sim 10^{55}$\,erg \citep{krieger2019}, an order of magnitude lower than that of M82 \citep{walter2002}.  Notably, the wholesale rearrangement of the magnetic field seen in the plane of M82 \citep{pattle2021a} is not replicated here: the average magnetic field angle in the centre of the galaxy is similar to the position angle of the major axis of the Nuclear Ring.  This is consistent with the behaviour seen at higher densities: \citet{lopezrodriguez2023}, observing the magnetic field in the central $\sim 10^{\prime\prime}$ around the galactic centre with ALMA, also found both that the magnetic field is preferentially aligned along the major axis of the Nuclear Ring, and that there is a field component aligned with the outflow within sites of active star formation.

\subsection{Spiral Arms}
\label{sec:discussion_arms}

Away from the galactic centre, we see clear qualitative trend for the magnetic field to run parallel to the spiral structure of the galaxy.  This behaviour is expected both from galactic dynamo models \citep{brandenburg2023} and from observations of other, non-starbursting, spiral galaxies \citep[e.g.,][]{scarrott1987,borlaff2021,lopezrodriguez2022}.  This suggests that away from the central outflow, as expected, we are observing polarization arising from cold dust in the galactic plane, and so tracing the galactic magnetic field.

The \textit{Clumpfind} emission peaks that we list in Table~\ref{tab:clumpfind} represent dust overdensities in the spiral arms of NGC 253 that are associated with sites of star formation (compare Figures~\ref{fig:clumps} and \ref{fig:sfr}).  However, it should be emphasised that we do not resolve individual giant molecular clouds (GMCs); with the typical size of a GMC being a few tens of pc and the largest being only $\sim 10^{2}$\,pc \citep{xie2024}, our 240\,pc beam integrates over entire clouds or cloud complexes, allowing us to observe only their net magnetic field properties.

\begin{figure}
    \centering
    \includegraphics[width=0.47\textwidth]{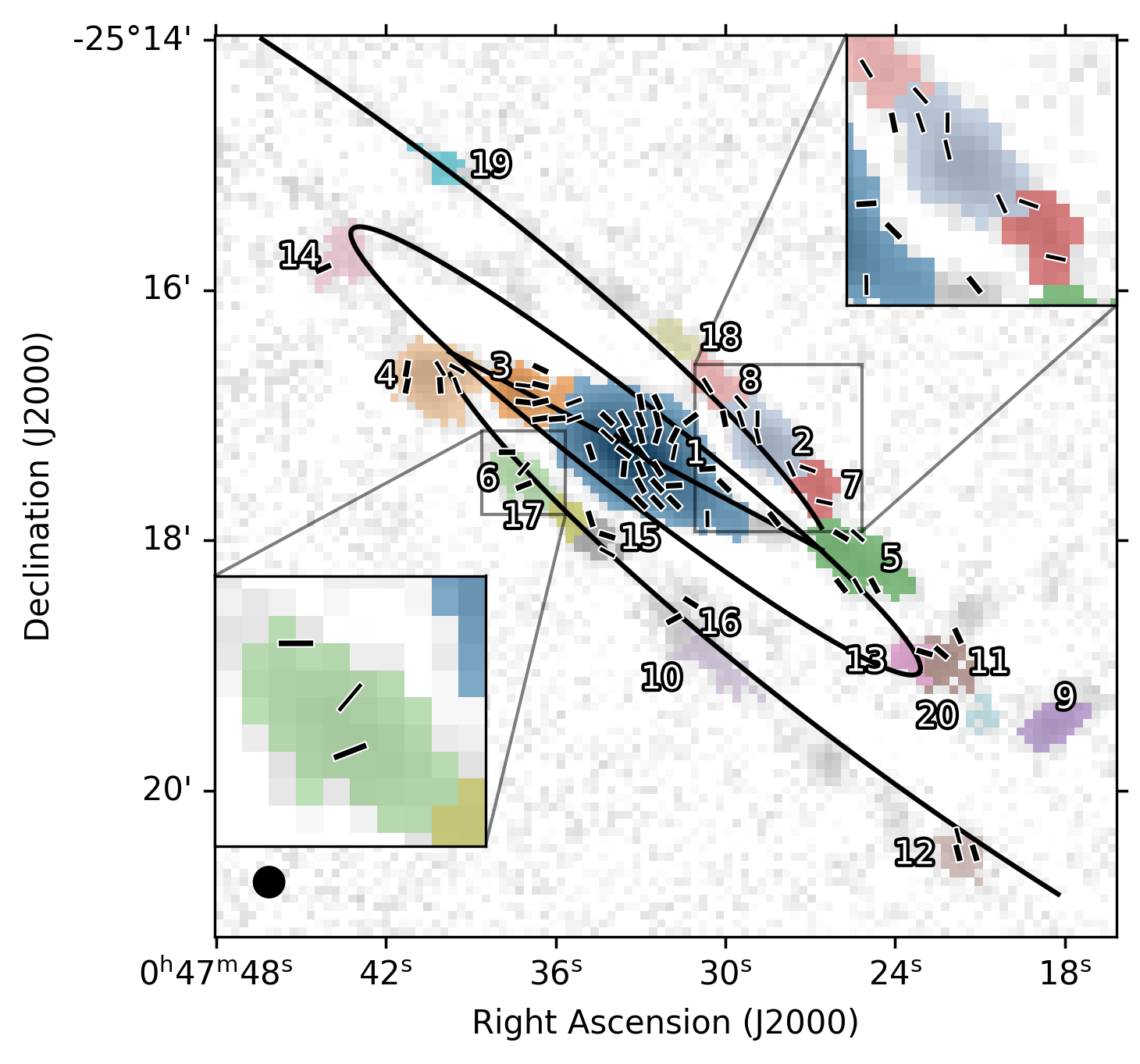}
    \caption{Magnetic field vectors and spiral structure model overplotted on \textit{Clumpfind} clumps.  Inset panels show zoomed-in images of Clumps 2 and 6, which respectively show depolarization and reorganisation of the magnetic field.}
    \label{fig:clumps_detail}
\end{figure}

In individual clumps within NGC 253, we see two behaviours which suggest that the magnetic field with GMC complexes is significantly distorted from the mean field behaviour.  These are (1) depolarization of 850\,$\mu$m emission associated with sites of star formation, and (2) large-scale reordering of the magnetic field from the local direction of the spiral structure.  Both of these behaviours are associated with the presence of stellar feedback.  For each of these behaviours we present a case study: Clumps 2 and 6 respectively, highlighted in Figure~\ref{fig:clumps_detail}, before discussing the quantification of these behaviours.

\subsubsection{Depolarization case study: Clump 2}
\label{sec:depol_cs}

Although the average magnetic field direction in Clump 2 is parallel to the spiral arm, it is notable for having no detections on the 850\,$\mu$m peak position (co-located with the peak in SFR).  Clump 2 is the third-brightest source that we detect, and we have achieved detections in significantly fainter material, suggesting that some property of either the magnetic field or the dust grains is preventing a detection.

Clump 2 is the most active star-forming region outside the galactic centre (see Figure~\ref{fig:sfr}).  There are four radio sources located within one JCMT beam of its central coordinates: \citet{ulvestad2000} Sources 4 (SNR), 6 (H\textsc{ii} region), 7 (SNR) and 8 (H\textsc{ii} region).  There is also an IR `hot spot' associated with the source \citep{kraemer2002} and an HCN peak \citep{beslic2024}.  As well as having an associated X-ray source \citep{colbert2004}, we also see an X-ray contour associated with Clump 2 in Figure~\ref{fig:xray}.

We thus hypothesise that within our 240\,pc beam we are integrating over a complex, and potentially significantly disordered, GMC-scale field in Clump 2, thereby preventing a detection of polarized emission on-peak.  This is consistent with our finding in Section~\ref{sec:pI} that the magnetic field direction becomes increasingly randomised as surface brightness increases.  This increase in apparent field disorder could be caused by turbulent fragmentation of the GMC cloud complex \citep{maclow2004}, and/or by large-scale reorganisation of the gas and dust in the region by stellar feedback \citep{geen2015}.

We note that Clumps 4, 5, 7 and 8 shows a similar behaviour, with no detection of polarized emission on peak.  In the cases of Clumps 4 and 5, their location, at the intersection (in projection) of the galactic bar, ring and arm, leads us to expect multiple ordered magnetic field components to be present along the line of sight.  However, it is notable that the depolarization in Clump 4 is associated again with a strong SFR peak, \citet{ulvestad2000} sources, and the presence of X-ray emission.  Clumps 7 and 8, located in a spiral arm, are also associated with \citet{ulvestad2000} sources, and in the case of Clump 8, with a peak in star formation rate.

\subsubsection{Depolarization: statistical consideration and interpretation}

We next attempt to put the behaviours discussed in Section~\ref{sec:depol_cs} on a statistical footing.  Figure~\ref{fig:sfr} suggests that there is an anticorrelation between POL-2 detections and both SFR and the presence of stellar/supernova feedback, as indicated by the presence of \citet{ulvestad2000} radio sources.  This is consistent with our interpretation that the large-scale galactic dynamo field becomes disordered or rearranged on small scales by star formation and feedback.  To quantify this finding, we tested the spatial correlation between POL-2 detections and \citet{ulvestad2000} sources using the \citet{chow2025} implementation of the \citet{fasano1987} (FF) multivariate Kolmogorov-Smirnov test\footnote{\url{https://github.com/wmpg/fasano-franceschini-test}}.  This implementation uses a permutation approach to significance testing, and so we ran the FF multiple times for each case considered in order to determine the range of $p$ values associated with the calculated FF test statistic.

We first compared two-dimensional distribution of the full set of POL-2 detections to the positions of the \citet{ulvestad2000} sources, and found a FF test statistic of 2.12, equivalent to a $p$ value in the range 0.002--0.004, indicating a $<0.5$\% probability that the two sets of positions are drawn from the same underlying distribution. However, this test included POL-2 vectors in the galactic centre, while \citet{ulvestad2000} lists only points in the galactic bar and arms, and so the test statistic may be artificially large.  We therefore repeated the test using only POL-2 detections outside the galactic centre, using the same mask as shown in Figure~\ref{fig:sfr}.  In this case, we found an FF test statistic of 1.49, equivalent to a $p$ value in the range 0.03--0.04, indicating a $<5$\% chance that the two sets of positions are drawn from the same underlying distribution.  We thus consider this anticorrelation statistically robust, and potentially indicative of the galactic-scale field becoming substantially disordered in sites of star formation and stellar feedback.

We note that this behaviour is consistent with analysis of HAWC+ observations of nearby galaxies \citep{borlaff2021, lopezrodriguez2022, borlaff2023, surgent2023}, and in synthetic observations of simulations \citep{martinalvarez2024}, in all of which depolarization and disordered magnetic fields are seen to be correlated with peaks in FIR emission, and so with star formation.  This behaviour is ascribed, as we do here, to field tangling due to physical processes associated with star formation activity operating on scales smaller than the telescope beam.

\begin{figure}
    \centering
    \includegraphics[width=0.47\textwidth]{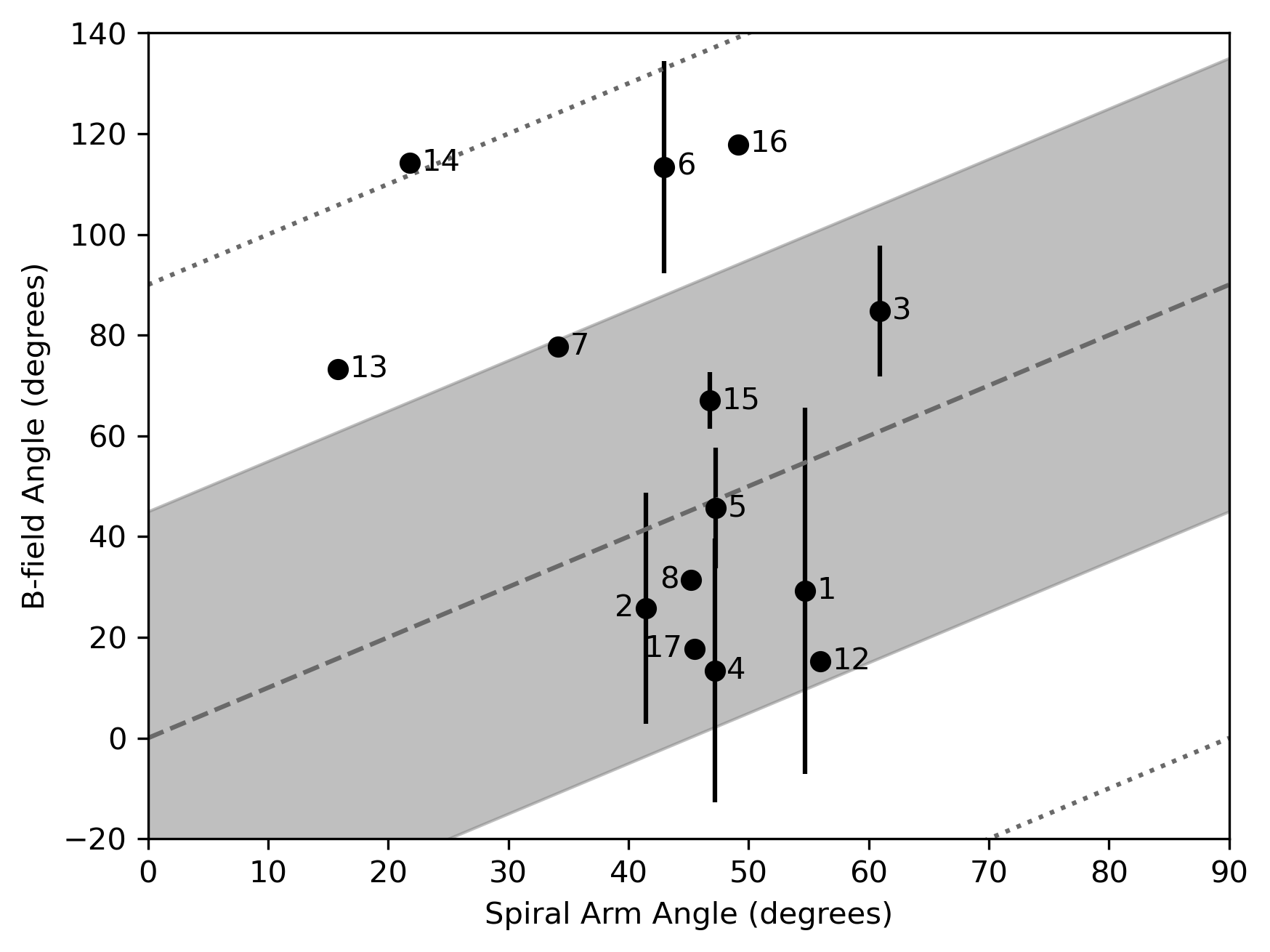}
    \caption{Comparison between magnetic field angle and spiral structure, for each \textit{Clumpfind} clump.  Dashed line indicates parallel alignment between magnetic field and spiral structure; dotted lines indicate perpendicular alignment.  Grey shaded area indicates alignments within 45$^{\circ}$ of parallel.  Clumps 7, 8, 13, 14, 16 and 17 each contain $< 3$ magnetic field vectors, and so are presented without error bars.  The error bars on Clump 12 are smaller than the data point.}
    \label{fig:spiral_clumps}
\end{figure}

\begin{figure*}
    \centering
    \includegraphics[width=\textwidth]{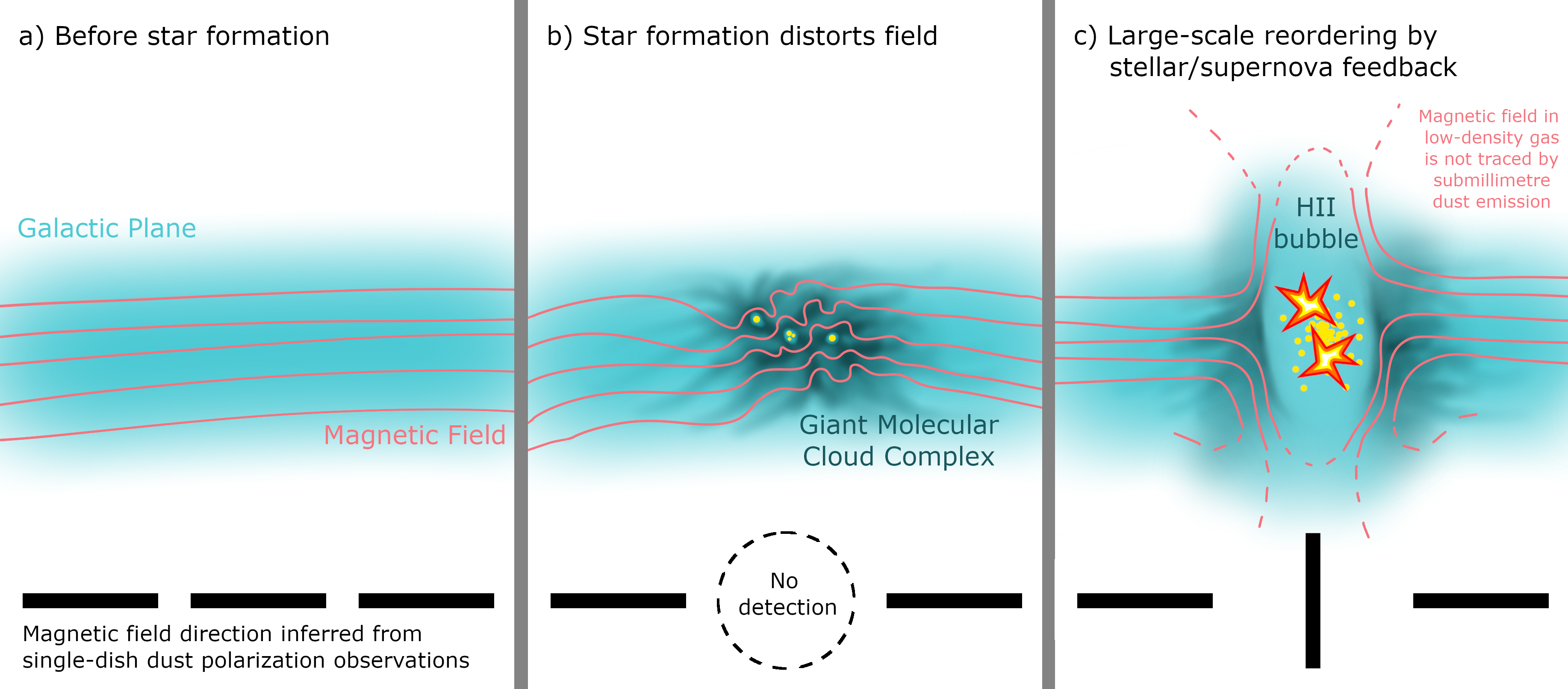}
    \caption{A cartoon illustrating our hypothesised evolutionary sequence for magnetic fields in GMCs in the spiral arms of NGC 253.  In panel (a), the unperturbed galactic dynamo magnetic field runs along the spiral arm of the galaxy.  In panel (b) the turbulent motions and gravitational collapse of a GMC, along with local radiative feedback from young stars, has significantly distorted the magnetic field, resulting in significant depolarization in the $\sim 10^{2}$\,pc beam of a single-dish telescope such as the JCMT.  In panel (c), a combination of radiative and supernova feedback has created a significant H\textsc{ii} bubble, which forms a chimney out of the plane of the galaxy.  The magnetic field in the dense gas in the plane of the galaxy, to which dust polarization observations are sensitive, has been significantly reordered by feedback to be perpendicular to the galactic plane.}
    \label{fig:cartoon}
\end{figure*}

\subsubsection{Reorganisation case study: Clump 6}

The most significant reordering of the galactic-scale magnetic field that we see is in Clump 6, in which the magnetic field is offset $70^{\circ}\pm 21^{\circ}$ from the local spiral arm direction.  This region is a site of active star formation, albeit currently less active than Clump 2: it has an associated radio source \citep[][Source 17]{ulvestad2000} and X-ray sources (\citealt{fabbiano1984} Source 7, and the transient \citealt{wik2014} Source 18), and is visible as an HCN peak \citep{beslic2024}.  We hypothesise that we are here observing a magnetic field that has been reordered to be preferentially perpendicular to the plane of the galaxy by significant stellar/supernova feedback, resulting from gas and dust being displaced by an expanding ionization (super)bubble driven by \citet{ulvestad2000} Source 17 and associated star formation in the region.  This rearrangement of the field in Clump 6, rather than depolarization seen in Clump 2, leads us to speculate that Clump 6 is more evolved than Clump 2.  This scenario is discussed further in Section~\ref{sec:discussion_all}, below.

We note that Source 6 lies along the same line of sight as the edge of the blue-shifted wing of the outflow from the galactic centre.  While it is possible that the magnetic field behaviour that we observe could be associated with the outflow, we consider this to be unlikely on three grounds: (1) even in the presence of a significant galactic outflow, POL-2 emission is strongly associated with cold and dense material in the plane of the galaxy \citep{pattle2021a}; if the emission we see were associated with the outflow, it would arise from material located $\sim 0.7$\,kpc above the plane of the galaxy. The dust scale height of NGC 253 is $\approx 0.15$\,kpc \citep{radovich2001}, and so sufficient dust emission being present at this height to create the observed signal is unlikely.  (2) Clump 6 is demonstrably a site of star formation, with which we expect significant dust continuum emission to be associated \citep[e.g.,][]{pattle2023}.  (3) Clump 15 is also co-located along the line of sight with the galactic outflow, and indeed overlaps further with the outflow than Clump 6, but shows no sign either of rearrangement of the field or of the presence of significant stellar/supernova feedback.

\subsubsection{Reorganisation: statistical consideration and interpretation}
\label{sec:discussion_reorg_stats}

The average angle between magnetic field direction and spiral structure direction is plotted in Figure~\ref{fig:spiral_clumps}.  Clump 6 is the only region containing more than one vector which shows a significant offset ($> 45^{\circ}$) from the direction of the local spiral structure in our 850\,$\mu$m data, although Clump 12 is also systematically offset from the local spiral arm direction. This could indicate some exceptional behaviour in Clump 6, or that Clump 6 represents a transient phase in the lifetime of a GMC complex: after the peak of star formation, but sufficiently affected by stellar and/or supernova feedback for a large-scale rearrangement of the magnetic field to have formed and to persist.

We note that a significant offset in magnetic field direction is also seen in HAWC+ 89\,$\mu$m observations of Clumps 2 and 3 (Figure~\ref{fig:hawc_data}).  In these two clumps - and particularly notably in Clump 2, which is depolarised in 850\,$\mu$m data, as discussed above -- the magnetic field traced by FIR emission is directed out of the plane, as seen in the submillimetre in Clump 6.  Both Clumps 2 and 3 are associated with \citet{ulvestad2000} sources.  This suggests that the magnetic fields in the hot dust traced by these FIR observations -- presumably heated by stellar feedback -- may be rearranged by at an earlier stage in the evolution of GMC complexes than is the cooler material traced by submillimetre dust emission.

\subsection{Synthesis and proposed evolutionary scenario}
\label{sec:discussion_all}

Figure~\ref{fig:cartoon} illustrates our proposed explanation for both the depolarization and the reorganisation of magnetic fields that we see in NGC 253.  We hypothesise that magnetic fields in GMCs undergo an evolutionary sequence from (a) a relatively ordered field parallel to the spiral arms in the pre-star-formation gas, to (b) a complex field caused by the interplay of turbulence, gravity and small-scale feedback in relatively young GMCs, to (c) a field that is reordered to be directed out of the plane of the galaxy as large-scale H\textsc{ii} region expansion reshapes the gas and dust of the spiral arm in which the H\textsc{ii} region resides.

In this hypothesis, we assume that ideal magnetohydrodynamic assumptions hold throughout, i.e. that the magnetic field is frozen into the gas, and so that if the gas and dust of the GMC complex is advected either by gravitational collapse or feedback expansion, the magnetic field will move with it \citep{mestel1956}.  This is a safe assumption when observing on GMC scales or larger; non-ideal effects are not expected to become significant except on very small ($\ll 1$\,pc) scales in the very densest regions within molecular clouds \citep{myers2018}. 

The reordering of magnetic fields in GMCs is well-attested \citep[][and refs. therein]{pattle2023a}; prior to the onset of star formation, magnetic fields typically evolve to become perpendicular to dense, self-gravitating filamentary structure \citep{soler2013, soler2017}, while after the onset of star formation, stellar feedback appears to reshape magnetic fields to typically run parallel to the boundaries of PDRs \citep[e.g.][]{pattle2018, khan2024}.  While the magnetic field is seen to be highly ordered in each of these cases, the increasing complexity, when averaged over the beam of a single-dish telescope, will result in the cancellation of signal to produce the `polarization hole' effect that we observe.

\citet{li2011}, observing GMCs in M33, found that their average magnetic field directions are preferentially aligned along the spiral arms of the galaxy.  This result is consistent with our own findings: \citet{li2011} observed at a linear resolution of $\sim 15$\,pc; the significant field complexity that they observe within GMCs in M33 would likely result in signal being integrated away in our 240\,pc beam.  Moreover, with an inclination of $\approx 54^{\circ}$ \citep{devaucouleurs1991}, M33 is significantly closer to being face-on than is the highly-inclined NGC 253; observations of the plane-of-sky magnetic field direction in M33 would therefore be significantly less sensitive to any component of the magnetic field perpendicular to the plane of the galaxy.

JWST observations have shown that feedback-driven (super)bubbles are ubiquitous in spiral galaxies, with diameters reaching $\gtrsim 1$\,kpc \citep{barnes2023}, and so we can expect reordering of magnetic fields in dense gas by such bubbles to be observable by the JCMT.  Our expectation is that the pressure gradient in the galactic plane will create ``chimneys'' in which ionized material can more easily escape out of the plane, and so we may expect flux-frozen magnetic field lines to be advected as this buoyant gas rises \citep{parker1966}.  Candidates for such chimneys have been observed in the Milky Way \citep[e.g.,][]{callaway2000,dawson2008,suad2022}, along with less dramatic supernova-feedback-driven reshaping of the local H\textsc{i} distribution \citep[e.g.,][]{soler2025}, although their associated magnetic fields have not yet been studied in detail.  Simulations of such chimneys which include magnetic fields suggest that magnetic fields will be enhanced by compression in the walls of such chimneys and that magnetic flux will be transported out of the plane \citet{kim2017}, although predictions for the direction, rather than the strength, of the magnetic field are not currently available. 

We note that in Figure~\ref{fig:cartoon}, we do not predict where the magnetic field lines in the low-density material above the plane of the galaxy connect.  Such supernova driven bubbles can stall and deposit their energy locally in the galactic arm; form a galactic fountain, returning material and energy to elsewhere in the plane of the galaxy; or can break out, connecting to the intergalactic medium \citep[e.g.,][]{fielding2018,orr2022}.  However, SCUBA-2/POL-2 observations are sensitive only to the coldest and densest material in the plane of the galaxy \citep{wardthompson2016,pattle2021a}, and so we expect to see only the magnetic field which is reordered in the walls of the chimney.

The behaviour that we hypothesise in the arms of NGC 253 likely forms a continuum with the more dramatic outflows that we see from starburst regions in the centres of NGC 253 (this work) and M82 \citep{pattle2021a}.  We note that in the centre of NGC 253 itself, the magnetic field shows a bimodal behaviour consistent with the evolutionary sequence that we posit for GMCs in the spiral arms, being oriented along the plane of the galaxy in lower-density regions, and out of the plane of the galaxy in sites of star formation \citep{lopezrodriguez2023}.

\subsubsection{Testing our hypothesised evolutionary sequence}

We have here hypothesised an evolutionary scenario to explain the variety of magnetic field morphologies that we observe in NGC 253.  We here consider how this hypothesis can be tested.

Most importantly, our hypothesis that Clump 6 is more evolved than Clump 2 must be tested, in order to validate our proposed evolutionary scenario.
Age estimates of young stellar clusters such as those forming in Clumps 2 and 6 can be made by measuring the equivalent widths (EWs) of the H$\alpha$, H$\beta$ or Br$\gamma$ lines; their EWs decrease as clusters age \citep{levesque2013}.  While H$\alpha$ emission has been mapped across the disc of NGC 253 \citep{congiu2025}, the high inclination angle of the galaxy means that emission from star-forming regions in both the nucleus and the arms is significantly extincted.  In extincted environments, Br$\gamma$ EW is the most suitable tracer of cluster age \citep{levesque2013}; indeed, Br$\gamma$ emission has previously been used to age star clusters in the nuclear starburst of NGC 253 \citep{kornei2009}.  Br$\gamma$ has a wavelength of 2.166\,$\mu$m; K-band follow-up of the clumps that we consider here is crucial to verify our hypothesis.  We will thus propose observations using the K-band Multi Object Spectrograph (KMOS) on the Very Large Telescope (VLT), or NIRCam on the James Webb Space Telescope (JWST).

We further note that the sample size in this work is modest, and emphasise the need both for more single-dish polarization observations of nearby galaxies and for higher-resolution observations of regions such as Clump 2 and Clump 6 in both NGC 253 and other nearby galaxies in order to test this hypothesis.  Such single-dish observations could be a science goal for the forthcoming PRobe far-Infrared Mission for Astrophysics (PRIMA; e.g. \citealt{maglione2025}), while high-resolution follow-up of these low-surface-brightness sources will likely require next-generation interferometers \citep{duartecabral2026}.

Additionally, predictions of magnetic field direction from MHD simulations of feedback chimneys, alongside radiative transfer post-processing of simulations to create synthetic observations against which our data can be compared, are crucial for testing the validity of our proposed scenario.

\section{Conclusions}
\label{sec:conclusions}

In this paper we present JCMT POL-2 850\,$\mu$m dust emission polarization observations of the nearby starburst galaxy NGC 253.  We find that the magnetic field that we observe, which principally arises from dense material located in the plane of the galaxy, typically traces the galactic spiral structure, in keeping with galactic dynamo models.  However, we also see significant deviations in magnetic field direction associated with the presence of stellar and/or supernova feedback.  The average magnetic field direction that we measure is $29^{\circ}$ E of N, with significant variations, compared to a galactic plane angle of $52^{\circ}$ E of N.

We find that polarization fraction is anticorrelated with total intensity, with a power-law index of $0.72\pm 0.03$.  We find that this depolarization is associated with high star formation rates and the presence of stellar/supernova feedback; using the Fasano-Franceschini multivariate Kolmogorov-Schmirnov test, we find that the positions of POL-2 detections and \citet{ulvestad2000} radio sources are not consistent with being drawn from the same underlying distribution.  This suggests that in sites of star formation and significant feedback, magnetic fields are tangled on scales smaller than the 240\,pc JCMT 850\,$\mu$m beam.

We identify peaks in 850\,$\mu$m intensity using the \textit{Clumpfind} algorithm.  The average magnetic field direction is typically preferentially aligned with the local spiral arm direction, albeit with depolarization on peak in a number of clumps.  However, in the starburst superwind above the centre of the galaxy, and in one star-forming clump in the spiral arms, we observe a magnetic field that has been systematically deviated from the direction of the spiral arm, consistent with being directed out of the plane of the galaxy.

We hypothesise that magnetic fields in the GMCs of NGC 253 are undergoing an evolutionary sequence from a relatively ordered field parallel to the spiral arms in the pre-star-formation gas, to a complex field caused by the interplay of turbulence, gravity and small-scale feedback in relatively young GMCs, to a field that is reordered to be directed out of the plane of the galaxy as large-scale H\textsc{ii} region expansion reshapes the gas and dust of the spiral arm to form a chimney.  In order to test this hypothesis, we need to verify the ages of the clouds; to do so in these heavily-extincted, highly-inclined clouds, we will propose to measure the equivalent widths of their Br$\gamma$ emission lines using instruments such as VLT/KMOS and JWST/NIRCam.  More generally, higher-resolution polarization observations of extragalactic GMCs, alongside predictions of magnetic field direction from MHD simulations of feedback chimneys, are necessary in order to test our proposed scenario.

\section*{Acknowledgements}

K.P. is a Royal Society University Research Fellow, supported by grant numbers URF\textbackslash R1\textbackslash 211322 and URF\textbackslash R\textbackslash 261015.

These observations were obtained by the James Clerk Maxwell Telescope, operated by the East Asian Observatory on behalf of The National Astronomical Observatory of Japan; Academia Sinica Institute of Astronomy and Astrophysics; the Korea Astronomy and Space Science Institute; the National Astronomical Research Institute of Thailand; Center for Astronomical Mega-Science (as well as the National Key R\&D Program of China with No. 2017YFA0402700). Additional funding support is provided by the Science and Technology Facilities Council of the United Kingdom and participating universities and organizations in the United Kingdom and Canada.

The Starlink software \citep{currie2014} is currently supported by the East Asian Observatory.  This work made use of Astropy (\url{http://www.astropy.org}), a community-developed core Python package and an ecosystem of tools and resources for astronomy \citep{astropy:2013, astropy:2018, astropy:2022}.

The authors wish to recognize and acknowledge the very significant cultural role and reverence that the summit of Maunakea has always had within the indigenous Hawaiian community.  We are most fortunate to have the opportunity to conduct observations from this mountain.

\section*{Data Availability}

The raw data used in this paper are available from the Canadian Astronomy Data Centre (CADC) archive under project codes M21BP024 and M22BP006.  The reduced data are available at [DOI TO BE INSERTED IN PROOF].



\bibliographystyle{mnras}

\begin{thebibliography}{}
\makeatletter
\relax
\def\mn@urlcharsother{\let\do\@makeother \do\$\do\&\do\#\do\^\do\_\do\%\do\~}
\def\mn@doi{\begingroup\mn@urlcharsother \@ifnextchar [ {\mn@doi@}
  {\mn@doi@[]}}
\def\mn@doi@[#1]#2{\def\@tempa{#1}\ifx\@tempa\@empty \href
  {http://dx.doi.org/#2} {doi:#2}\else \href {http://dx.doi.org/#2} {#1}\fi
  \endgroup}
\def\mn@eprint#1#2{\mn@eprint@#1:#2::\@nil}
\def\mn@eprint@arXiv#1{\href {http://arxiv.org/abs/#1} {{\tt arXiv:#1}}}
\def\mn@eprint@dblp#1{\href {http://dblp.uni-trier.de/rec/bibtex/#1.xml}
  {dblp:#1}}
\def\mn@eprint@#1:#2:#3:#4\@nil{\def\@tempa {#1}\def\@tempb {#2}\def\@tempc
  {#3}\ifx \@tempc \@empty \let \@tempc \@tempb \let \@tempb \@tempa \fi \ifx
  \@tempb \@empty \def\@tempb {arXiv}\fi \@ifundefined
  {mn@eprint@\@tempb}{\@tempb:\@tempc}{\expandafter \expandafter \csname
  mn@eprint@\@tempb\endcsname \expandafter{\@tempc}}}

\bibitem[\protect\citeauthoryear{{Adebahr}, {Krause}, {Klein}, {Heald}  \&
  {Dettmar}}{{Adebahr} et~al.}{2017}]{adebahr2017}
{Adebahr} B.,  {Krause} M.,  {Klein} U.,  {Heald} G.,   {Dettmar} R.~J.,  2017,
  \mn@doi [\aap] {10.1051/0004-6361/201629616}, \href
  {https://ui.adsabs.harvard.edu/abs/2017A&A...608A..29A} {608, A29}

\bibitem[\protect\citeauthoryear{{Alves}, {Frau}, {Girart}, {Franco}, {Santos}
  \& {Wiesemeyer}}{{Alves} et~al.}{2014}]{alves2014}
{Alves} F.~O.,  {Frau} P.,  {Girart} J.~M.,  {Franco} G.~A.~P.,  {Santos}
  F.~P.,   {Wiesemeyer} H.,  2014, \mn@doi [\aap]
  {10.1051/0004-6361/201424678}, \href
  {https://ui.adsabs.harvard.edu/abs/2014A&A...569L...1A} {569, L1}

\bibitem[\protect\citeauthoryear{{Andersson}, {Lazarian}  \&
  {Vaillancourt}}{{Andersson} et~al.}{2015}]{andersson2015}
{Andersson} B.~G.,  {Lazarian} A.,   {Vaillancourt} J.~E.,  2015, \mn@doi
  [\araa] {10.1146/annurev-astro-082214-122414}, \href
  {https://ui.adsabs.harvard.edu/abs/2015ARA&A..53..501A} {53, 501}

\bibitem[\protect\citeauthoryear{{Astropy Collaboration} et~al.,}{{Astropy
  Collaboration} et~al.}{2013}]{astropy:2013}
{Astropy Collaboration} et~al., 2013, \mn@doi [\aap]
  {10.1051/0004-6361/201322068}, \href
  {http://adsabs.harvard.edu/abs/2013A%26A...558A..33A} {558, A33}

\bibitem[\protect\citeauthoryear{{Astropy Collaboration} et~al.,}{{Astropy
  Collaboration} et~al.}{2018}]{astropy:2018}
{Astropy Collaboration} et~al., 2018, \mn@doi [\aj] {10.3847/1538-3881/aabc4f},
  \href {https://ui.adsabs.harvard.edu/abs/2018AJ....156..123A} {156, 123}

\bibitem[\protect\citeauthoryear{{Astropy Collaboration} et~al.,}{{Astropy
  Collaboration} et~al.}{2022}]{astropy:2022}
{Astropy Collaboration} et~al., 2022, \mn@doi [apj] {10.3847/1538-4357/ac7c74},
  \href {https://ui.adsabs.harvard.edu/abs/2022ApJ...935..167A} {935, 167}

\bibitem[\protect\citeauthoryear{{Barnes} et~al.,}{{Barnes}
  et~al.}{2023}]{barnes2023}
{Barnes} A.~T.,  et~al., 2023, \mn@doi [\apjl] {10.3847/2041-8213/aca7b9},
  \href {https://ui.adsabs.harvard.edu/abs/2023ApJ...944L..22B} {944, L22}

\bibitem[\protect\citeauthoryear{{Belfiori} et~al.,}{{Belfiori}
  et~al.}{2025}]{belfiori2025}
{Belfiori} D.,  et~al., 2025, \mn@doi [\aap] {10.1051/0004-6361/202451615},
  \href {https://ui.adsabs.harvard.edu/abs/2025A&A...697A..24B} {697, A24}

\bibitem[\protect\citeauthoryear{{Beslic} et~al.,}{{Beslic}
  et~al.}{2024}]{beslic2024}
{Beslic} I.,  et~al., 2024, \mn@doi [arXiv e-prints]
  {10.48550/arXiv.2403.13751}, \href
  {https://ui.adsabs.harvard.edu/abs/2024arXiv240313751B} {p. arXiv:2403.13751}

\bibitem[\protect\citeauthoryear{{Bolatto} et~al.,}{{Bolatto}
  et~al.}{2013}]{bolatto2013}
{Bolatto} A.~D.,  et~al., 2013, \mn@doi [\nat] {10.1038/nature12351}, \href
  {https://ui.adsabs.harvard.edu/abs/2013Natur.499..450B} {499, 450}

\bibitem[\protect\citeauthoryear{{Borlaff} et~al.,}{{Borlaff}
  et~al.}{2021}]{borlaff2021}
{Borlaff} A.~S.,  et~al., 2021, \mn@doi [\apj] {10.3847/1538-4357/ac16d7},
  \href {https://ui.adsabs.harvard.edu/abs/2021ApJ...921..128B} {921, 128}

\bibitem[\protect\citeauthoryear{{Borlaff} et~al.,}{{Borlaff}
  et~al.}{2023}]{borlaff2023}
{Borlaff} A.~S.,  et~al., 2023, \mn@doi [\apj] {10.3847/1538-4357/acd934},
  \href {https://ui.adsabs.harvard.edu/abs/2023ApJ...952....4B} {952, 4}

\bibitem[\protect\citeauthoryear{{Brandenburg} \& {Ntormousi}}{{Brandenburg} \&
  {Ntormousi}}{2023}]{brandenburg2023}
{Brandenburg} A.,  {Ntormousi} E.,  2023, \mn@doi [\araa]
  {10.1146/annurev-astro-071221-052807}, \href
  {https://ui.adsabs.harvard.edu/abs/2023ARA&A..61..561B} {61, 561}

\bibitem[\protect\citeauthoryear{{Callaway}, {Savage}, {Benjamin}, {Haffner}
  \& {Tufte}}{{Callaway} et~al.}{2000}]{callaway2000}
{Callaway} M.~B.,  {Savage} B.~D.,  {Benjamin} R.~A.,  {Haffner} L.~M.,
  {Tufte} S.~L.,  2000, \mn@doi [\apj] {10.1086/308601}, \href
  {https://ui.adsabs.harvard.edu/abs/2000ApJ...532..943C} {532, 943}

\bibitem[\protect\citeauthoryear{{Chapin}, {Berry}, {Gibb}, {Jenness}, {Scott},
  {Tilanus}, {Economou}  \& {Holland}}{{Chapin} et~al.}{2013}]{chapin2013}
{Chapin} E.~L.,  {Berry} D.~S.,  {Gibb} A.~G.,  {Jenness} T.,  {Scott} D.,
  {Tilanus} R. P.~J.,  {Economou} F.,   {Holland} W.~S.,  2013, \mn@doi
  [\mnras] {10.1093/mnras/stt052}, \href
  {https://ui.adsabs.harvard.edu/abs/2013MNRAS.430.2545C} {430, 2545}

\bibitem[\protect\citeauthoryear{{Chow} \& {Brown}}{{Chow} \&
  {Brown}}{2025}]{chow2025}
{Chow} I.,  {Brown} P.~G.,  2025, \mn@doi [\icarus]
  {10.1016/j.icarus.2024.116444}, \href
  {https://ui.adsabs.harvard.edu/abs/2025Icar..42916444C} {429, 116444}

\bibitem[\protect\citeauthoryear{{Clark} et~al.,}{{Clark}
  et~al.}{2018}]{clark2018}
{Clark} C.~J.~R.,  et~al., 2018, \mn@doi [\aap] {10.1051/0004-6361/201731419},
  \href {https://ui.adsabs.harvard.edu/abs/2018A&A...609A..37C} {609, A37}

\bibitem[\protect\citeauthoryear{{Colbert}, {Heckman}, {Ptak}, {Strickland}  \&
  {Weaver}}{{Colbert} et~al.}{2004}]{colbert2004}
{Colbert} E. J.~M.,  {Heckman} T.~M.,  {Ptak} A.~F.,  {Strickland} D.~K.,
  {Weaver} K.~A.,  2004, \mn@doi [\apj] {10.1086/380899}, \href
  {https://ui.adsabs.harvard.edu/abs/2004ApJ...602..231C} {602, 231}

\bibitem[\protect\citeauthoryear{{Congiu} et~al.,}{{Congiu}
  et~al.}{2025}]{congiu2025}
{Congiu} E.,  et~al., 2025, \mn@doi [\aap] {10.1051/0004-6361/202554144}, \href
  {https://ui.adsabs.harvard.edu/abs/2025A&A...700A.125C} {700, A125}

\bibitem[\protect\citeauthoryear{{Cronin} et~al.,}{{Cronin}
  et~al.}{2025}]{cronin2025}
{Cronin} S.~A.,  et~al., 2025, \mn@doi [\apj] {10.3847/1538-4357/add738}, \href
  {https://ui.adsabs.harvard.edu/abs/2025ApJ...987...92C} {987, 92}

\bibitem[\protect\citeauthoryear{{Currie}, {Berry}, {Jenness}, {Gibb}, {Bell}
  \& {Draper}}{{Currie} et~al.}{2014}]{currie2014}
{Currie} M.~J.,  {Berry} D.~S.,  {Jenness} T.,  {Gibb} A.~G.,  {Bell} G.~S.,
  {Draper} P.~W.,  2014, in {Manset} N.,  {Forshay} P.,  eds,  Astronomical
  Society of the Pacific Conference Series Vol. 485, Astronomical Data Analysis
  Software and Systems XXIII. p.~391

\bibitem[\protect\citeauthoryear{{Davidge}}{{Davidge}}{2010}]{davidge2010}
{Davidge} T.~J.,  2010, \mn@doi [\apj] {10.1088/0004-637X/725/1/1342}, \href
  {https://ui.adsabs.harvard.edu/abs/2010ApJ...725.1342D} {725, 1342}

\bibitem[\protect\citeauthoryear{{Dawson}, {Mizuno}, {Onishi},
  {McClure-Griffiths}  \& {Fukui}}{{Dawson} et~al.}{2008}]{dawson2008}
{Dawson} J.~R.,  {Mizuno} N.,  {Onishi} T.,  {McClure-Griffiths} N.~M.,
  {Fukui} Y.,  2008, \mn@doi [\mnras] {10.1111/j.1365-2966.2008.13152.x}, \href
  {https://ui.adsabs.harvard.edu/abs/2008MNRAS.387...31D} {387, 31}

\bibitem[\protect\citeauthoryear{{Dempsey} et~al.,}{{Dempsey}
  et~al.}{2013}]{dempsey2013}
{Dempsey} J.~T.,  et~al., 2013, \mn@doi [\mnras] {10.1093/mnras/stt090}, \href
  {https://ui.adsabs.harvard.edu/abs/2013MNRAS.430.2534D} {430, 2534}

\bibitem[\protect\citeauthoryear{{Duarte-Cabral} et~al.,}{{Duarte-Cabral}
  et~al.}{2026}]{duartecabral2026}
{Duarte-Cabral} A.,  et~al., 2026, \mn@doi [arXiv e-prints]
  {10.48550/arXiv.2610.21661}, \href
  {https://ui.adsabs.harvard.edu/abs/2026arXiv261021661D} {p. arXiv:2610.21661}

\bibitem[\protect\citeauthoryear{{Fabbiano} \& {Trinchieri}}{{Fabbiano} \&
  {Trinchieri}}{1984}]{fabbiano1984}
{Fabbiano} G.,  {Trinchieri} G.,  1984, \mn@doi [\apj] {10.1086/162623}, \href
  {https://ui.adsabs.harvard.edu/abs/1984ApJ...286..491F} {286, 491}

\bibitem[\protect\citeauthoryear{{Fasano} \& {Franceschini}}{{Fasano} \&
  {Franceschini}}{1987}]{fasano1987}
{Fasano} G.,  {Franceschini} A.,  1987, \mn@doi [\mnras]
  {10.1093/mnras/225.1.155}, \href
  {https://ui.adsabs.harvard.edu/abs/1987MNRAS.225..155F} {225, 155}

\bibitem[\protect\citeauthoryear{{Fielding}, {Quataert}  \&
  {Martizzi}}{{Fielding} et~al.}{2018}]{fielding2018}
{Fielding} D.,  {Quataert} E.,   {Martizzi} D.,  2018, \mn@doi [\mnras]
  {10.1093/mnras/sty2466}, \href
  {https://ui.adsabs.harvard.edu/abs/2018MNRAS.481.3325F} {481, 3325}

\bibitem[\protect\citeauthoryear{{Ford} et~al.,}{{Ford}
  et~al.}{2013}]{ford2013}
{Ford} G.~P.,  et~al., 2013, \mn@doi [\apj] {10.1088/0004-637X/769/1/55}, \href
  {https://ui.adsabs.harvard.edu/abs/2013ApJ...769...55F} {769, 55}

\bibitem[\protect\citeauthoryear{{Friberg}, {Bastien}, {Berry}, {Savini},
  {Graves}  \& {Pattle}}{{Friberg} et~al.}{2016}]{friberg2016}
{Friberg} P.,  {Bastien} P.,  {Berry} D.,  {Savini} G.,  {Graves} S.~F.,
  {Pattle} K.,  2016, in {Holland} W.~S.,  {Zmuidzinas} J.,  eds,  Society of
  Photo-Optical Instrumentation Engineers (SPIE) Conference Series Vol. 9914,
  Millimeter, Submillimeter, and Far-Infrared Detectors and Instrumentation for
  Astronomy VIII. p. 991403, \mn@doi{10.1117/12.2231943}

\bibitem[\protect\citeauthoryear{{Geen}, {Hennebelle}, {Tremblin}  \&
  {Rosdahl}}{{Geen} et~al.}{2015}]{geen2015}
{Geen} S.,  {Hennebelle} P.,  {Tremblin} P.,   {Rosdahl} J.,  2015, \mn@doi
  [\mnras] {10.1093/mnras/stv2272}, \href
  {https://ui.adsabs.harvard.edu/abs/2015MNRAS.454.4484G} {454, 4484}

\bibitem[\protect\citeauthoryear{{Heesen}, {Beck}, {Krause}  \&
  {Dettmar}}{{Heesen} et~al.}{2011}]{heesen2011}
{Heesen} V.,  {Beck} R.,  {Krause} M.,   {Dettmar} R.-J.,  2011, \mn@doi [\aap]
  {10.1051/0004-6361/201117618}, \href
  {https://ui.adsabs.harvard.edu/abs/2011A&A...535A..79H} {535, A79}

\bibitem[\protect\citeauthoryear{{Hildebrand}}{{Hildebrand}}{1983}]{hildebrand1983}
{Hildebrand} R.~H.,  1983, \qjras, \href
  {https://ui.adsabs.harvard.edu/abs/1983QJRAS..24..267H} {24, 267}

\bibitem[\protect\citeauthoryear{{Holland} et~al.,}{{Holland}
  et~al.}{2013}]{holland2013}
{Holland} W.~S.,  et~al., 2013, \mn@doi [\mnras] {10.1093/mnras/sts612}, \href
  {https://ui.adsabs.harvard.edu/abs/2013MNRAS.430.2513H} {430, 2513}

\bibitem[\protect\citeauthoryear{{Iodice}, {Arnaboldi}, {Rejkuba}, {Neeser},
  {Greggio}, {Gonzalez}, {Irwin}  \& {Emerson}}{{Iodice}
  et~al.}{2014}]{iodice2014}
{Iodice} E.,  {Arnaboldi} M.,  {Rejkuba} M.,  {Neeser} M.~J.,  {Greggio} L.,
  {Gonzalez} O.~A.,  {Irwin} M.,   {Emerson} J.~P.,  2014, \mn@doi [\aap]
  {10.1051/0004-6361/201423480}, \href
  {https://ui.adsabs.harvard.edu/abs/2014A&A...567A..86I} {567, A86}

\bibitem[\protect\citeauthoryear{{Jones}, {Bagley}, {Krejny}, {Andersson}  \&
  {Bastien}}{{Jones} et~al.}{2015}]{jones2015}
{Jones} T.~J.,  {Bagley} M.,  {Krejny} M.,  {Andersson} B.~G.,   {Bastien} P.,
  2015, \mn@doi [\aj] {10.1088/0004-6256/149/1/31}, \href
  {https://ui.adsabs.harvard.edu/abs/2015AJ....149...31J} {149, 31}

\bibitem[\protect\citeauthoryear{{Jones} et~al.,}{{Jones}
  et~al.}{2019}]{jones2019}
{Jones} T.~J.,  et~al., 2019, \mn@doi [\apjl] {10.3847/2041-8213/aaf8b9}, \href
  {https://ui.adsabs.harvard.edu/abs/2019ApJ...870L...9J} {870, L9}

\bibitem[\protect\citeauthoryear{{Jones} et~al.,}{{Jones}
  et~al.}{2020}]{jones2020}
{Jones} T.~J.,  et~al., 2020, \mn@doi [\aj] {10.3847/1538-3881/abada8}, \href
  {https://ui.adsabs.harvard.edu/abs/2020AJ....160..167J} {160, 167}

\bibitem[\protect\citeauthoryear{{Karoly} et~al.,}{{Karoly}
  et~al.}{2023}]{karoly2023}
{Karoly} J.,  et~al., 2023, \mn@doi [\apj] {10.3847/1538-4357/acd6f2}, \href
  {https://ui.adsabs.harvard.edu/abs/2023ApJ...952...29K} {952, 29}

\bibitem[\protect\citeauthoryear{{Khan}, {Pattle}  \& {Graves}}{{Khan}
  et~al.}{2024}]{khan2024}
{Khan} Z.~A.,  {Pattle} K.,   {Graves} S.~F.,  2024, \mn@doi [\mnras]
  {10.1093/mnras/stae2350}, \href
  {https://ui.adsabs.harvard.edu/abs/2024MNRAS.535..107K} {535, 107}

\bibitem[\protect\citeauthoryear{{Kim} \& {Ostriker}}{{Kim} \&
  {Ostriker}}{2017}]{kim2017}
{Kim} C.-G.,  {Ostriker} E.~C.,  2017, \mn@doi [\apj]
  {10.3847/1538-4357/aa8599}, \href
  {https://ui.adsabs.harvard.edu/abs/2017ApJ...846..133K} {846, 133}

\bibitem[\protect\citeauthoryear{{Kornei} \& {McCrady}}{{Kornei} \&
  {McCrady}}{2009}]{kornei2009}
{Kornei} K.~A.,  {McCrady} N.,  2009, \mn@doi [\apj]
  {10.1088/0004-637X/697/2/1180}, \href
  {https://ui.adsabs.harvard.edu/abs/2009ApJ...697.1180K} {697, 1180}

\bibitem[\protect\citeauthoryear{{Kraemer}, {Price}, {Mizuno}  \&
  {Carey}}{{Kraemer} et~al.}{2002}]{kraemer2002}
{Kraemer} K.~E.,  {Price} S.~D.,  {Mizuno} D.~R.,   {Carey} S.~J.,  2002,
  \mn@doi [\aj] {10.1086/344079}, \href
  {https://ui.adsabs.harvard.edu/abs/2002AJ....124.2990K} {124, 2990}

\bibitem[\protect\citeauthoryear{{Krieger} et~al.,}{{Krieger}
  et~al.}{2019}]{krieger2019}
{Krieger} N.,  et~al., 2019, \mn@doi [\apj] {10.3847/1538-4357/ab2d9c}, \href
  {https://ui.adsabs.harvard.edu/abs/2019ApJ...881...43K} {881, 43}

\bibitem[\protect\citeauthoryear{{Leroy}, {Walter}, {Brinks}, {Bigiel}, {de
  Blok}, {Madore}  \& {Thornley}}{{Leroy} et~al.}{2008}]{leroy2008}
{Leroy} A.~K.,  {Walter} F.,  {Brinks} E.,  {Bigiel} F.,  {de Blok} W.~J.~G.,
  {Madore} B.,   {Thornley} M.~D.,  2008, \mn@doi [\aj]
  {10.1088/0004-6256/136/6/2782}, \href
  {https://ui.adsabs.harvard.edu/abs/2008AJ....136.2782L} {136, 2782}

\bibitem[\protect\citeauthoryear{{Levesque} \& {Leitherer}}{{Levesque} \&
  {Leitherer}}{2013}]{levesque2013}
{Levesque} E.~M.,  {Leitherer} C.,  2013, \mn@doi [\apj]
  {10.1088/0004-637X/779/2/170}, \href
  {https://ui.adsabs.harvard.edu/abs/2013ApJ...779..170L} {779, 170}

\bibitem[\protect\citeauthoryear{{Li} \& {Henning}}{{Li} \&
  {Henning}}{2011}]{li2011}
{Li} H.-B.,  {Henning} T.,  2011, \mn@doi [\nat] {10.1038/nature10551}, \href
  {https://ui.adsabs.harvard.edu/abs/2011Natur.479..499L} {479, 499}

\bibitem[\protect\citeauthoryear{{Lopez-Rodriguez}}{{Lopez-Rodriguez}}{2023}]{lopezrodriguez2023}
{Lopez-Rodriguez} E.,  2023, \mn@doi [\apj] {10.3847/1538-4357/ace110}, \href
  {https://ui.adsabs.harvard.edu/abs/2023ApJ...953..113L} {953, 113}

\bibitem[\protect\citeauthoryear{{Lopez-Rodriguez}, {Guerra}, {Asgari-Targhi}
  \& {Schmelz}}{{Lopez-Rodriguez} et~al.}{2021}]{lopezrodriguez2021}
{Lopez-Rodriguez} E.,  {Guerra} J.,  {Asgari-Targhi} M.,   {Schmelz} J.~T.,
  2021, arXiv e-prints, \href
  {https://ui.adsabs.harvard.edu/abs/2021arXiv210203362L} {p. arXiv:2102.03362}

\bibitem[\protect\citeauthoryear{{Lopez-Rodriguez} et~al.,}{{Lopez-Rodriguez}
  et~al.}{2022}]{lopezrodriguez2022}
{Lopez-Rodriguez} E.,  et~al., 2022, \mn@doi [\apj] {10.3847/1538-4357/ac7f9d},
  \href {https://ui.adsabs.harvard.edu/abs/2022ApJ...936...92L} {936, 92}

\bibitem[\protect\citeauthoryear{{Lopez}, {Lopez}, {Nguyen}, {Thompson},
  {Mathur}, {Bolatto}, {Vulic}  \& {Sardone}}{{Lopez} et~al.}{2023}]{lopez2023}
{Lopez} S.,  {Lopez} L.~A.,  {Nguyen} D.~D.,  {Thompson} T.~A.,  {Mathur} S.,
  {Bolatto} A.~D.,  {Vulic} N.,   {Sardone} A.,  2023, \mn@doi [\apj]
  {10.3847/1538-4357/aca65e}, \href
  {https://ui.adsabs.harvard.edu/abs/2023ApJ...942..108L} {942, 108}

\bibitem[\protect\citeauthoryear{{Mac Low} \& {Klessen}}{{Mac Low} \&
  {Klessen}}{2004}]{maclow2004}
{Mac Low} M.-M.,  {Klessen} R.~S.,  2004, \mn@doi [Reviews of Modern Physics]
  {10.1103/RevModPhys.76.125}, \href
  {https://ui.adsabs.harvard.edu/abs/2004RvMP...76..125M} {76, 125}

\bibitem[\protect\citeauthoryear{{Maglione}, {Martin-Alvarez},
  {Lopez-Rodriguez}, {Clark}  \& {Karpovich}}{{Maglione}
  et~al.}{2025}]{maglione2025}
{Maglione} D.,  {Martin-Alvarez} S.,  {Lopez-Rodriguez} E.,  {Clark} S.~E.,
  {Karpovich} K.,  2025, \mn@doi [Journal of Astronomical Telescopes,
  Instruments, and Systems] {10.1117/1.JATIS.11.3.031620}, \href
  {https://ui.adsabs.harvard.edu/abs/2025JATIS..11c1620M} {11, 031620}

\bibitem[\protect\citeauthoryear{{Mairs} et~al.,}{{Mairs}
  et~al.}{2021}]{mairs2021}
{Mairs} S.,  et~al., 2021, \mn@doi [\aj] {10.3847/1538-3881/ac18bf}, \href
  {https://ui.adsabs.harvard.edu/abs/2021AJ....162..191M} {162, 191}

\bibitem[\protect\citeauthoryear{{Martin-Alvarez} et~al.,}{{Martin-Alvarez}
  et~al.}{2024}]{martinalvarez2024}
{Martin-Alvarez} S.,  et~al., 2024, \mn@doi [\apj] {10.3847/1538-4357/ad2e9e},
  \href {https://ui.adsabs.harvard.edu/abs/2024ApJ...966...43M} {966, 43}

\bibitem[\protect\citeauthoryear{{Mestel} \& {Spitzer}}{{Mestel} \&
  {Spitzer}}{1956}]{mestel1956}
{Mestel} L.,  {Spitzer} Jr. L.,  1956, \mn@doi [\mnras]
  {10.1093/mnras/116.5.503}, \href
  {https://ui.adsabs.harvard.edu/abs/1956MNRAS.116..503M} {116, 503}

\bibitem[\protect\citeauthoryear{{Montier}, {Plaszczynski}, {Levrier},
  {Tristram}, {Alina}, {Ristorcelli}, {Bernard}  \& {Guillet}}{{Montier}
  et~al.}{2015}]{montier2015}
{Montier} L.,  {Plaszczynski} S.,  {Levrier} F.,  {Tristram} M.,  {Alina} D.,
  {Ristorcelli} I.,  {Bernard} J.~P.,   {Guillet} V.,  2015, \mn@doi [\aap]
  {10.1051/0004-6361/201424451}, \href
  {https://ui.adsabs.harvard.edu/abs/2015A&A...574A.136M} {574, A136}

\bibitem[\protect\citeauthoryear{{Myers}, {Basu}  \& {Auddy}}{{Myers}
  et~al.}{2018}]{myers2018}
{Myers} P.~C.,  {Basu} S.,   {Auddy} S.,  2018, \mn@doi [\apj]
  {10.3847/1538-4357/aae695}, \href
  {https://ui.adsabs.harvard.edu/abs/2018ApJ...868...51M} {868, 51}

\bibitem[\protect\citeauthoryear{{Naghizadeh-Khouei} \&
  {Clarke}}{{Naghizadeh-Khouei} \& {Clarke}}{1993}]{naghizadehkhouei1993}
{Naghizadeh-Khouei} J.,  {Clarke} D.,  1993, \aap, \href
  {https://ui.adsabs.harvard.edu/abs/1993A&A...274..968N} {274, 968}

\bibitem[\protect\citeauthoryear{{Newman}, {McQuinn}, {Skillman}, {Boyer},
  {Cohen}, {Dolphin}  \& {Telford}}{{Newman} et~al.}{2024}]{newman2024}
{Newman} M. J.~B.,  {McQuinn} K. B.~W.,  {Skillman} E.~D.,  {Boyer} M.~L.,
  {Cohen} R.~E.,  {Dolphin} A.~E.,   {Telford} O.~G.,  2024, \mn@doi [\apj]
  {10.3847/1538-4357/ad306d}, \href
  {https://ui.adsabs.harvard.edu/abs/2024ApJ...966..175N} {966, 175}

\bibitem[\protect\citeauthoryear{{Okamoto} et~al.,}{{Okamoto}
  et~al.}{2024}]{okamoto2024}
{Okamoto} S.,  et~al., 2024, \mn@doi [\apjl] {10.3847/2041-8213/ad4358}, \href
  {https://ui.adsabs.harvard.edu/abs/2024ApJ...967L..24O} {967, L24}

\bibitem[\protect\citeauthoryear{{Orr}, {Fielding}, {Hayward}  \&
  {Burkhart}}{{Orr} et~al.}{2022}]{orr2022}
{Orr} M.~E.,  {Fielding} D.~B.,  {Hayward} C.~C.,   {Burkhart} B.,  2022,
  \mn@doi [\apjl] {10.3847/2041-8213/ac479f}, \href
  {https://ui.adsabs.harvard.edu/abs/2022ApJ...924L..28O} {924, L28}

\bibitem[\protect\citeauthoryear{{Ott}, {Weiss}, {Henkel}  \& {Walter}}{{Ott}
  et~al.}{2005}]{ott2005}
{Ott} J.,  {Weiss} A.,  {Henkel} C.,   {Walter} F.,  2005, \mn@doi [\apj]
  {10.1086/431661}, \href
  {https://ui.adsabs.harvard.edu/abs/2005ApJ...629..767O} {629, 767}

\bibitem[\protect\citeauthoryear{{Parker}}{{Parker}}{1966}]{parker1966}
{Parker} E.~N.,  1966, \mn@doi [\apj] {10.1086/148828}, \href
  {https://ui.adsabs.harvard.edu/abs/1966ApJ...145..811P} {145, 811}

\bibitem[\protect\citeauthoryear{{Pattle} et~al.,}{{Pattle}
  et~al.}{2018}]{pattle2018}
{Pattle} K.,  et~al., 2018, \mn@doi [\apjl] {10.3847/2041-8213/aac771}, \href
  {https://ui.adsabs.harvard.edu/abs/2018ApJ...860L...6P} {860, L6}

\bibitem[\protect\citeauthoryear{{Pattle} et~al.,}{{Pattle}
  et~al.}{2019}]{pattle2019a}
{Pattle} K.,  et~al., 2019, \mn@doi [\apj] {10.3847/1538-4357/ab286f}, \href
  {https://ui.adsabs.harvard.edu/abs/2019ApJ...880...27P} {880, 27}

\bibitem[\protect\citeauthoryear{{Pattle}, {Gear}, {Redman}, {Smith}  \&
  {Greaves}}{{Pattle} et~al.}{2021a}]{pattle2021a}
{Pattle} K.,  {Gear} W.,  {Redman} M.,  {Smith} M. W.~L.,   {Greaves} J.,
  2021a, \mn@doi [\mnras] {10.1093/mnras/stab1300}, \href
  {https://ui.adsabs.harvard.edu/abs/2021MNRAS.505..684P} {505, 684}

\bibitem[\protect\citeauthoryear{{Pattle} et~al.,}{{Pattle}
  et~al.}{2021b}]{pattle2021}
{Pattle} K.,  et~al., 2021b, \mn@doi [\apj] {10.3847/1538-4357/abcc6c}, \href
  {https://ui.adsabs.harvard.edu/abs/2021ApJ...907...88P} {907, 88}

\bibitem[\protect\citeauthoryear{{Pattle}, {Gear}  \& {Wilson}}{{Pattle}
  et~al.}{2023a}]{pattle2023}
{Pattle} K.,  {Gear} W.,   {Wilson} C.~D.,  2023a, \mn@doi [\mnras]
  {10.1093/mnras/stad652}, \href
  {https://ui.adsabs.harvard.edu/abs/2023MNRAS.522.2339P} {522, 2339}

\bibitem[\protect\citeauthoryear{{Pattle}, {Fissel}, {Tahani}, {Liu}  \&
  {Ntormousi}}{{Pattle} et~al.}{2023b}]{pattle2023a}
{Pattle} K.,  {Fissel} L.,  {Tahani} M.,  {Liu} T.,   {Ntormousi} E.,  2023b,
  in {Inutsuka} S.,  {Aikawa} Y.,  {Muto} T.,  {Tomida} K.,   {Tamura} M.,
  eds,  Astronomical Society of the Pacific Conference Series Vol. 534,
  Protostars and Planets VII. p.~193 (\mn@eprint {arXiv} {2203.11179}),
  \mn@doi{10.48550/arXiv.2203.11179}

\bibitem[\protect\citeauthoryear{{Planck Collaboration} et~al.,}{{Planck
  Collaboration} et~al.}{2020}]{planck2020}
{Planck Collaboration} et~al., 2020, \mn@doi [\aap]
  {10.1051/0004-6361/201833885}, \href
  {https://ui.adsabs.harvard.edu/abs/2020A&A...641A..12P} {641, A12}

\bibitem[\protect\citeauthoryear{{Plaszczynski}, {Montier}, {Levrier}  \&
  {Tristram}}{{Plaszczynski} et~al.}{2014}]{plaszczynski2014}
{Plaszczynski} S.,  {Montier} L.,  {Levrier} F.,   {Tristram} M.,  2014,
  \mn@doi [\mnras] {10.1093/mnras/stu270}, \href
  {https://ui.adsabs.harvard.edu/abs/2014MNRAS.439.4048P} {439, 4048}

\bibitem[\protect\citeauthoryear{{Radovich}, {Kahanp{\"a}{\"a}}  \&
  {Lemke}}{{Radovich} et~al.}{2001}]{radovich2001}
{Radovich} M.,  {Kahanp{\"a}{\"a}} J.,   {Lemke} D.,  2001, \mn@doi [\aap]
  {10.1051/0004-6361:20011102}, \href
  {https://ui.adsabs.harvard.edu/abs/2001A&A...377...73R} {377, 73}

\bibitem[\protect\citeauthoryear{{Rekola}, {Richer}, {McCall}, {Valtonen},
  {Kotilainen}  \& {Flynn}}{{Rekola} et~al.}{2005}]{rekola2005}
{Rekola} R.,  {Richer} M.~G.,  {McCall} M.~L.,  {Valtonen} M.~J.,  {Kotilainen}
  J.~K.,   {Flynn} C.,  2005, \mn@doi [\mnras]
  {10.1111/j.1365-2966.2005.09166.x}, \href
  {https://ui.adsabs.harvard.edu/abs/2005MNRAS.361..330R} {361, 330}

\bibitem[\protect\citeauthoryear{{Rieder} \& {Teyssier}}{{Rieder} \&
  {Teyssier}}{2016}]{rieder2016}
{Rieder} M.,  {Teyssier} R.,  2016, \mn@doi [\mnras] {10.1093/mnras/stv2985},
  \href {https://ui.adsabs.harvard.edu/abs/2016MNRAS.457.1722R} {457, 1722}

\bibitem[\protect\citeauthoryear{{Rieder} \& {Teyssier}}{{Rieder} \&
  {Teyssier}}{2017}]{rieder2017}
{Rieder} M.,  {Teyssier} R.,  2017, \mn@doi [\mnras] {10.1093/mnras/stx1670},
  \href {https://ui.adsabs.harvard.edu/abs/2017MNRAS.471.2674R} {471, 2674}

\bibitem[\protect\citeauthoryear{{Sadavoy} et~al.,}{{Sadavoy}
  et~al.}{2013}]{sadavoy2013}
{Sadavoy} S.~I.,  et~al., 2013, \mn@doi [\apj] {10.1088/0004-637X/767/2/126},
  \href {https://ui.adsabs.harvard.edu/abs/2013ApJ...767..126S} {767, 126}

\bibitem[\protect\citeauthoryear{{Scarrott}, {Ward-Thompson}  \&
  {Warren-Smith}}{{Scarrott} et~al.}{1987}]{scarrott1987}
{Scarrott} S.~M.,  {Ward-Thompson} D.,   {Warren-Smith} R.~F.,  1987, \mn@doi
  [\mnras] {10.1093/mnras/224.2.299}, \href
  {https://ui.adsabs.harvard.edu/abs/1987MNRAS.224..299S} {224, 299}

\bibitem[\protect\citeauthoryear{{Serkowski}}{{Serkowski}}{1962}]{serkowski1962}
{Serkowski} K.,  1962, \mn@doi [Advances in Astronomy and Astrophysics]
  {10.1016/B978-1-4831-9919-1.50009-1}, \href
  {https://ui.adsabs.harvard.edu/abs/1962AdA&A...1..289S} {1, 289}

\bibitem[\protect\citeauthoryear{{Soler} \& {Hennebelle}}{{Soler} \&
  {Hennebelle}}{2017}]{soler2017}
{Soler} J.~D.,  {Hennebelle} P.,  2017, \mn@doi [\aap]
  {10.1051/0004-6361/201731049}, \href
  {https://ui.adsabs.harvard.edu/abs/2017A&A...607A...2S} {607, A2}

\bibitem[\protect\citeauthoryear{{Soler}, {Hennebelle}, {Martin},
  {Miville-Desch{\^e}nes}, {Netterfield}  \& {Fissel}}{{Soler}
  et~al.}{2013}]{soler2013}
{Soler} J.~D.,  {Hennebelle} P.,  {Martin} P.~G.,  {Miville-Desch{\^e}nes}
  M.-A.,  {Netterfield} C.~B.,   {Fissel} L.~M.,  2013, \mn@doi [\apj]
  {10.1088/0004-637X/774/2/128}, \href
  {https://ui.adsabs.harvard.edu/abs/2013ApJ...774..128S} {774, 128}

\bibitem[\protect\citeauthoryear{{Soler} et~al.,}{{Soler}
  et~al.}{2025}]{soler2025}
{Soler} J.~D.,  et~al., 2025, \mn@doi [\aap] {10.1051/0004-6361/202453022},
  \href {https://ui.adsabs.harvard.edu/abs/2025A&A...695A.222S} {695, A222}

\bibitem[\protect\citeauthoryear{{Suad}, {Molina Lera}  \&
  {Cichowolski}}{{Suad} et~al.}{2022}]{suad2022}
{Suad} L.~A.,  {Molina Lera} J.~A.,   {Cichowolski} S.,  2022, \mn@doi [\aap]
  {10.1051/0004-6361/202243942}, \href
  {https://ui.adsabs.harvard.edu/abs/2022A&A...668A..44S} {668, A44}

\bibitem[\protect\citeauthoryear{{Surgent}, {Lopez-Rodriguez}  \&
  {Clark}}{{Surgent} et~al.}{2023}]{surgent2023}
{Surgent} W.~J.,  {Lopez-Rodriguez} E.,   {Clark} S.~E.,  2023, \mn@doi [\apj]
  {10.3847/1538-4357/ace4c0}, \href
  {https://ui.adsabs.harvard.edu/abs/2023ApJ...954...53S} {954, 53}

\bibitem[\protect\citeauthoryear{{Thompson}, {Quataert}, {Waxman}, {Murray}  \&
  {Martin}}{{Thompson} et~al.}{2006}]{thompson2006}
{Thompson} T.~A.,  {Quataert} E.,  {Waxman} E.,  {Murray} N.,   {Martin} C.~L.,
   2006, \mn@doi [\apj] {10.1086/504035}, \href
  {https://ui.adsabs.harvard.edu/abs/2006ApJ...645..186T} {645, 186}

\bibitem[\protect\citeauthoryear{{Ulvestad}}{{Ulvestad}}{2000}]{ulvestad2000}
{Ulvestad} J.~S.,  2000, \mn@doi [\aj] {10.1086/301428}, \href
  {https://ui.adsabs.harvard.edu/abs/2000AJ....120..278U} {120, 278}

\bibitem[\protect\citeauthoryear{{Walter}, {Weiss}  \& {Scoville}}{{Walter}
  et~al.}{2002}]{walter2002}
{Walter} F.,  {Weiss} A.,   {Scoville} N.,  2002, \mn@doi [\apjl]
  {10.1086/345287}, \href
  {https://ui.adsabs.harvard.edu/abs/2002ApJ...580L..21W} {580, L21}

\bibitem[\protect\citeauthoryear{{Ward-Thompson} et~al.,}{{Ward-Thompson}
  et~al.}{2016}]{wardthompson2016}
{Ward-Thompson} D.,  et~al., 2016, \mn@doi [\mnras] {10.1093/mnras/stw1978},
  \href {https://ui.adsabs.harvard.edu/abs/2016MNRAS.463.1008W} {463, 1008}

\bibitem[\protect\citeauthoryear{{Whittet}, {Hough}, {Lazarian}  \&
  {Hoang}}{{Whittet} et~al.}{2008}]{whittet2008}
{Whittet} D.~C.~B.,  {Hough} J.~H.,  {Lazarian} A.,   {Hoang} T.,  2008,
  \mn@doi [\apj] {10.1086/525040}, \href
  {https://ui.adsabs.harvard.edu/abs/2008ApJ...674..304W} {674, 304}

\bibitem[\protect\citeauthoryear{{Wik} et~al.,}{{Wik} et~al.}{2014}]{wik2014}
{Wik} D.~R.,  et~al., 2014, \mn@doi [\apj] {10.1088/0004-637X/797/2/79}, \href
  {https://ui.adsabs.harvard.edu/abs/2014ApJ...797...79W} {797, 79}

\bibitem[\protect\citeauthoryear{{Williams}, {de Geus}  \& {Blitz}}{{Williams}
  et~al.}{1994}]{williams1994}
{Williams} J.~P.,  {de Geus} E.~J.,   {Blitz} L.,  1994, \mn@doi [\apj]
  {10.1086/174279}, \href
  {https://ui.adsabs.harvard.edu/abs/1994ApJ...428..693W} {428, 693}

\bibitem[\protect\citeauthoryear{{Xie}, {Li}  \& {Chen}}{{Xie}
  et~al.}{2024}]{xie2024}
{Xie} Y.-H.,  {Li} G.-X.,   {Chen} B.-Q.,  2024, \mn@doi [\apj]
  {10.3847/1538-4357/ad7378}, \href
  {https://ui.adsabs.harvard.edu/abs/2024ApJ...975...39X} {975, 39}

\bibitem[\protect\citeauthoryear{{de Vaucouleurs}, {de Vaucouleurs}, {Corwin},
  {Buta}, {Paturel}  \& {Fouque}}{{de Vaucouleurs}
  et~al.}{1991}]{devaucouleurs1991}
{de Vaucouleurs} G.,  {de Vaucouleurs} A.,  {Corwin} Jr. H.~G.,  {Buta} R.~J.,
  {Paturel} G.,   {Fouque} P.,  1991, {Third Reference Catalogue of Bright
  Galaxies}

\makeatother
\end{thebibliography}




\appendix

\section{Comparison with HAWC+ and radio data}
\label{sec:appendix}

\begin{figure*}
    \includegraphics[width=0.8\textwidth]{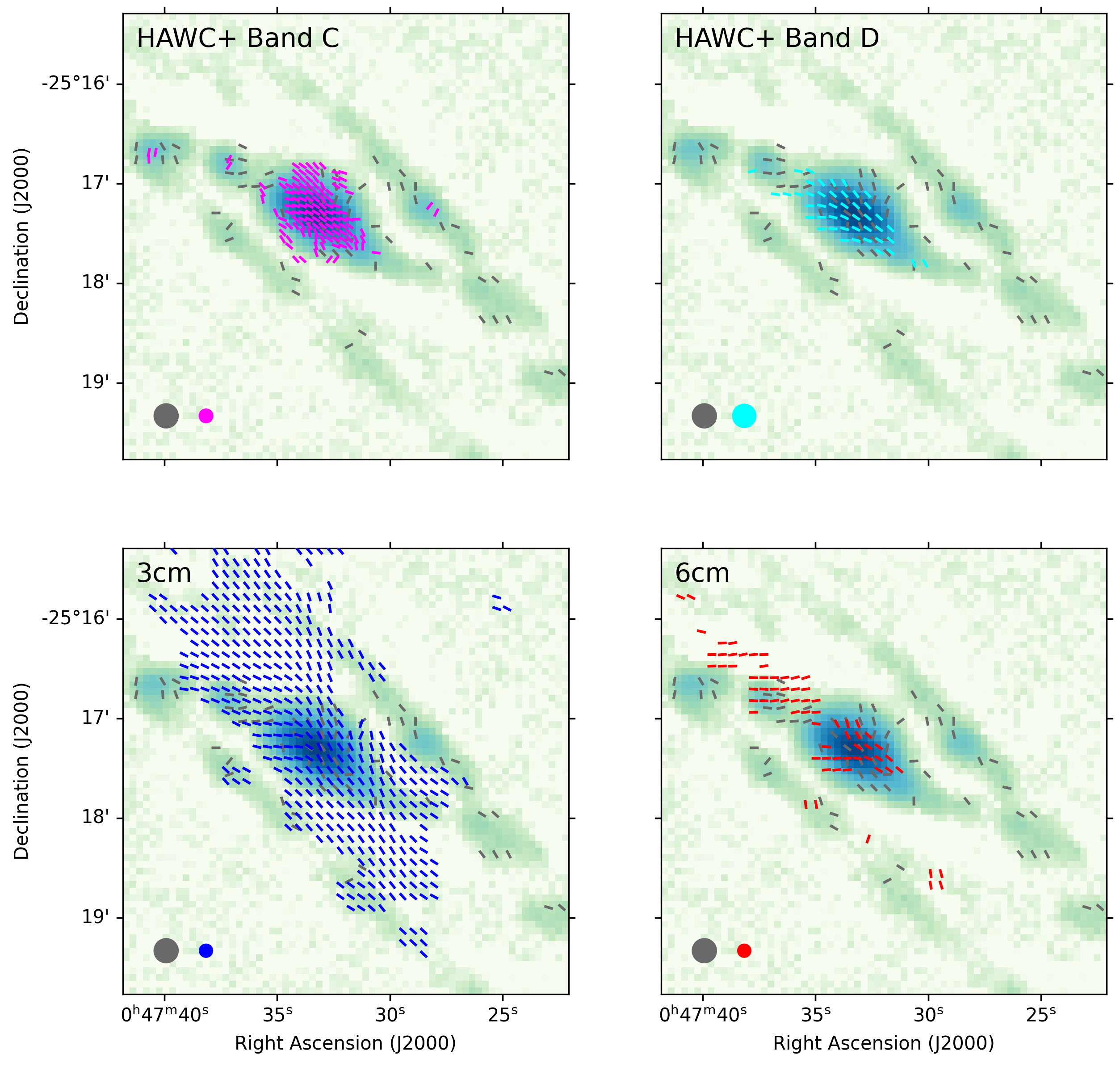}
    \caption{Magnetic field vectors at other wavelengths, overplotted on POL-2 850\,$\mu$m Stokes $I$ emission.  \textit{Top left:} HAWC+ Band C (89 $\mu$m); \textit{top right:} HAWC+ Band D (154 $\mu$m); \textit{bottom left:} VLA 3\,cm; \textit{bottom right:} VLA 6\,cm.  The beam sizes of each instrument are shown in the lower left-hand corner of each panel.  POL-2 magnetic field vectors (this work) are shown in grey.}
    \label{fig:hawc_data}
\end{figure*}

\begin{figure*}
    \includegraphics[width=0.8\textwidth]{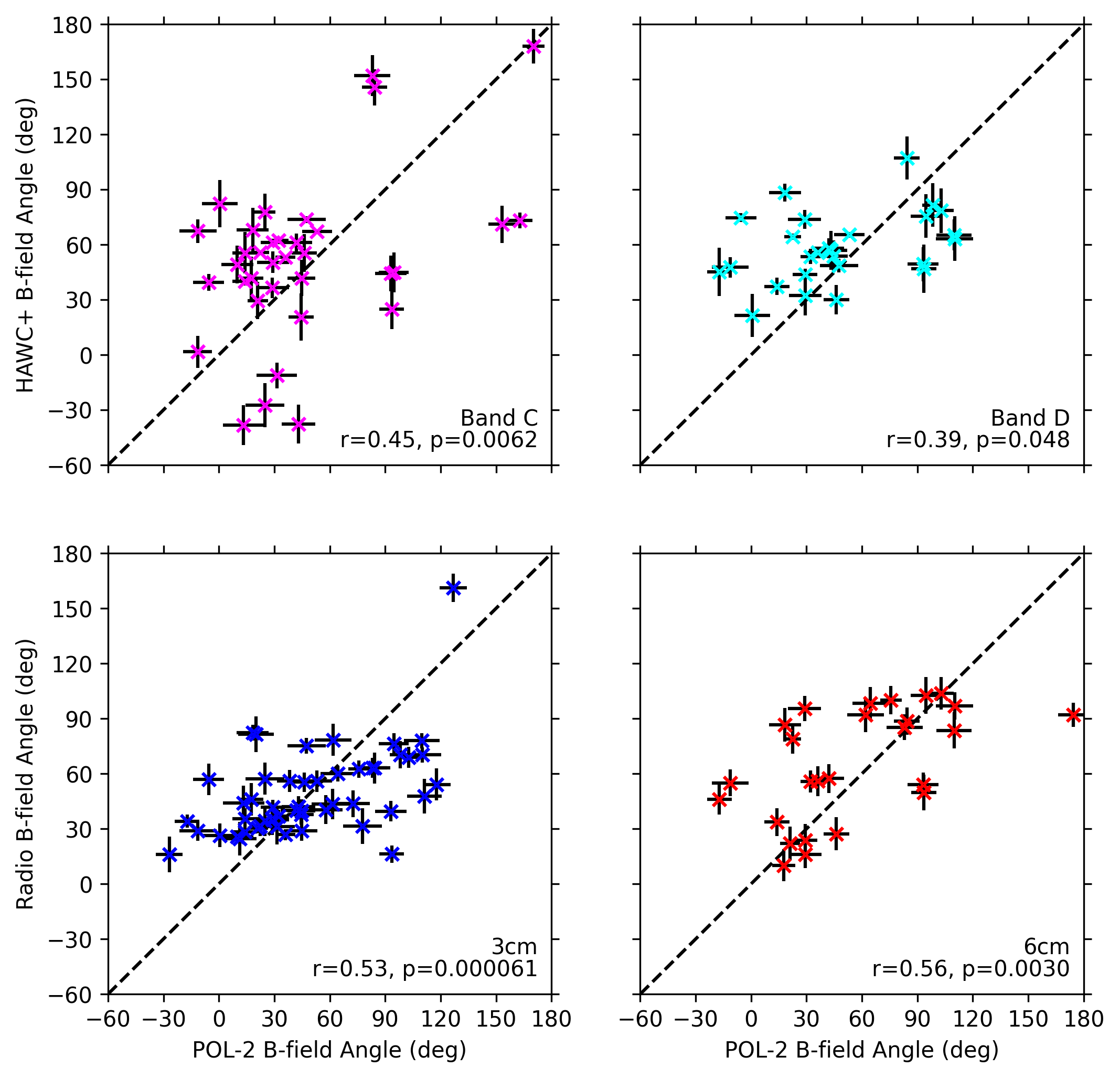}
    \caption{Comparison of magnetic field position angles between POL-2 and other wavelengths, for positions matched within 1 JCMT beam.  \textit{Top left:} HAWC+ Band C (89 $\mu$m) cross-match; \textit{top right:} HAWC+ Band D (154 $\mu$m) cross-match; \textit{bottom left:} VLA 3\,cm cross-match; \textit{bottom right:} VLA 6\,cm cross-match.  Dashed line shows the 1:1 relation.  Position angles are usually defined in the range $0^{\circ}\leq\theta_{B}<180^{\circ}$, with the difference in angle defined as $\theta_{B,\rm POL-2} - \theta_{B, \rm other}$.  However, since the maximum allowed absolute difference between a pair of angles is 90$^{\circ}$, and the range $-90^{\circ}$ -- $0^{\circ}$ is identical to the range $90^{\circ}$ -- $180^{\circ}$, when the difference in angle is $>90^{\circ}$ we shift $\theta_{B,\rm POL-2}$ by $-180^{\circ}$ and when the difference in angle is $< -90^{\circ}$ we shift $\theta_{B, \rm other}$ by  $-180^{\circ}$ to ensure that the correlation coefficients that we present are calculated for angle differences in the appropriate range.}
    \label{fig:hawc_crossmatch}
\end{figure*}

HAWC+ Band C and D dust polarization vectors \citep{jones2019,lopezrodriguez2022}, and VLA 3\,cm and 6\,cm synchrotron polarization vectors \citep{heesen2011}, are shown in Figure~\ref{fig:hawc_data}. These data were obtained from the SALSA survey archive\footnote{\url{http://galmagfields.com}}.

NGC253 was observed in HAWC+ Bands A (53\,$\mu$m), C (89\,$\mu$m) and D (154\,$\mu$m).  The detection in Band A is minimal, so these data are not shown. Bands C and D show good detections across the nuclear region and bar of NGC 253.  The vectors shown in Figure~\ref{fig:hawc_data} are selected to have $p/\delta p > 2$.  This is broadly in keeping with the SALSA survey convention of using $PI/\delta PI > 2$, although we note that $p/\delta p$ is a more stringent criterion than is $PI/\delta PI$ \citep{serkowski1962}.

Histograms of angle distribution are shown in Figure~\ref{fig:polar} in the main body of the text.  Bands C and D and the 3\,cm synchrotron data all have strong peaks at $\simeq 45-50^{\circ}$ E of N.  It appears that these data trace a magnetic field which is well-aligned with the major axis of the Nuclear Ring and the Outer Ring (both 52$^{\circ}$ E of N).  The mean magnetic field angle that we measure with POL-2 is $29^{\circ}$, albeit with substantial variations, suggesting that the 850\,$\mu$m emission is tracing both the galactic dynamo field and a component of the magnetic field associated with the outflow from the central starburst, as discussed in Section~\ref{sec:discussion_centre}.  The 6\,cm sychrotron vectors are less well-correlated with the plane of the galaxy, but appear to be primarily associated with both the galactic nucleus and the galactic bar, the latter of which has a significantly shallower position angle than do the two rings (we adopt 62$^{\circ}$ for the bar, while \citealt{iodice2014} choose an even shallower angle of 72$^{\circ}$), and so also appears to be tracing a galactic dynamo field.

We cross-matched our POL-2 vector catalogue with the HAWC+ and VLA vectors, in each case accepting closest matches within the POL-2 beam, which is in each case the larger beam of the pair. 
The magnetic field position angles of the cross-matched vectors are shown in Figure~\ref{fig:hawc_crossmatch}.  There is a reasonable correlation between the position angles measured by POL-2 and HAWC+, and between angles measured by POL-2 and VLA, with in each case a Spearman's rank correlation coefficient $p$ value $<0.05$.  Despite the discrepancies between these different wavelengths discussed above, all four wavelengths appear to broadly trace the same galactic magnetic field. 


\bsp	
\label{lastpage}
\end{document}